\documentclass[a4paper,11pt]{article}
\pdfoutput=1 
\usepackage{jheppub} 

\usepackage{amsmath,amssymb,amsfonts,cases}
\usepackage{amsthm}
\usepackage{enumitem}
\usepackage{adjustbox}
\usepackage{graphicx,caption,subcaption}
\usepackage[T1]{fontenc} 
\usepackage{enumitem}  
\usepackage{empheq}
\usepackage{dsfont}
\usepackage[mathscr]{euscript}
\usepackage{dcolumn}
\usepackage{bm}
\usepackage{bbm}
\usepackage{slashed}
\usepackage{wrapfig}
\usepackage{mathtools}
\usepackage{tikz}
\usepackage{hyperref}
\usepackage{wasysym}

\newcommand{\vol}[1]{\text{vol}(#1)~}
\newcommand{\beq}{\begin{equation}}
\newcommand{\eeq}{\end{equation}}
\newcommand{\ba}{\begin{align}}
\newcommand{\ea}{\end{align}}
\newcommand{\bea}{\begin{eqnarray}}
\newcommand{\eea}{\end{eqnarray}}
\newcommand{\bi}{\begin{itemize}}
\newcommand{\ei}{\end{itemize}}
\newcommand{\ben}{\begin{enumerate}}
\newcommand{\een}{\end{enumerate}}

\renewcommand{\a}{\alpha}

\newcommand{\e}{\epsilon}

\renewcommand{\k}{\kappa}

\newcommand{\n}{\nu}

\renewcommand{\r}{\rho}

\def\a{\alpha}

\def\k{\kappa}             

\def\n{\nu}

\def\r{\rho}                                     

\def\F{\Phi}

\newcommand{\pd}{\partial}

\newcommand{\eps}{\epsilon}

\newcommand\eq[1]{eq.~(\ref{eq:#1})}

\newcommand{\sn}[1]{section~\ref{sec:#1}}
\newcommand{\fig}[1]{figure~\ref{fig:#1}}
\newcommand{\app}[1]{appendix~\ref{app:#1}}

\newcommand\mf[1]{{\mathfrak{#1}}}
\newcommand\mc[1]{{\mathcal{#1}}}

\newcommand\CA{{\mc{A}}}
\newcommand\CB{{\mc{B}}}
\newcommand\CC{{\mc{C}}}

\newcommand\CI{{\mc{I}}}
\newcommand\CJ{{\mc{J}}}

\newcommand\CM{{\mc{M}}}
\newcommand\CN{{\mc{N}}}

\newcommand\CR{{\mc{R}}}
\newcommand\CS{{\mc{S}}}

\newcommand\CZ{{\mc{Z}}}

\newcommand\bcc{{a_{\Sigma}}}

\definecolor{cardinal}{rgb}{0.6,0,0}
\definecolor{darkgreen}{rgb}{0,0.5,0}
\definecolor{golden}{rgb}{0.92, 0.7, 0}
\definecolor{midnight}{rgb}{0, 0, 0.5}
\definecolor{darkblue}{rgb}{0.2, 0, 0.8}

\def\shrug{\texttt{\raisebox{0.75em}{\char`\_}\char`\\\char`\_\kern-0.5ex(\kern-0.25ex\raisebox{0.25ex}{\rotatebox{45}{\raisebox{-.75ex}"\kern-1.5ex\rotatebox{-90})}}\kern-0.5ex)\kern-0.5ex\char`\_/\raisebox{0.75em}{\char`\_}}}

\begin{document}
\title{\LARGE Holographic Weyl Anomalies for 4d Defects in 6d SCFTs}
\author[a]{Pietro Capuozzo}
\author[b]{\!\!, John Estes}
\author[c]{\!\!, Brandon Robinson}
\author[a]{\!\!, and Benjamin Suzzoni}
\emailAdd{p.capuozzo@soton.ac.uk}
\emailAdd{estesj@oldwestbury.edu}
\emailAdd{brandon.robinson@mib.infn.it}
\emailAdd{b.suzzoni@soton.ac.uk}
\affiliation[a]{STAG Research Centre, University of Southampton, Southampton, SO17 1BJ, UK}
\affiliation[b]{Department of Chemistry and Physics, SUNY Old Westbury, Old Westbury, NY, United States}
\affiliation[c]{INFN, sezione di Milano-Bicocca,
Piazza della Scienza 3, I-20126 Milano, Italy}
\abstract{
In this note, we study $1/4$- and $1/2$-BPS co-dimension two superconformal defects in the $6d$ $\mathcal{N}=(2,0)$ $A_{N-1}$ SCFT at large $N$ using their holographic descriptions as solutions of $11d$ supergravity.  In this regime, we are able to compute the defect contribution to the sphere entanglement entropy and the change in the stress-energy tensor one-point function due to the presence of the defect using holography.  From these quantities, we are then able to unambiguously compute the values for two of the twenty-nine total Weyl anomaly coefficients that characterize $4d$ conformal defects in six and higher dimensions. We are able to demonstrate the consistency of the supergravity description of the defect theories with the average null energy condition on the field theory side. For each class of defects that we consider, we also show that the A-type Weyl anomaly coefficient is non-negative. Lastly, we uncover and  resolve a discrepancy between the on-shell action of the $7d$ $1/4$-BPS domain wall solutions and that of their $11d$ uplift.
}

\arxivnumber{2310.17447}

\maketitle
\tableofcontents

\section{Introduction}

Knowing the spectrum of local operators in a given quantum field theory (QFT) is insufficient to uniquely specify it in field theory space \cite{Douglas:2010ic}, and so operators with non-trivial extension along submanifolds embedded in the background spacetime (`defects') play an important role in classifying QFTs \cite{Aharony:2013hda}.  However, the way that the presence of these defects affects, say, correlation functions of local operators depends on the dimension $d$ and geometry of the background manifold $\CM_d$, the co-dimension $d-\mf{d}$ and embedding of the $\mf{d}$-dimensional defect submanifold $\Sigma_{\mf{d}}$, and the couplings between ambient and defect degrees of freedom\footnote{See \cite{Andrei:2018die} for a recent review of defects of various (co-)dimension in QFTs.}.  Thus, it is crucial to characterize allowable defects in a given theory and precisely determine how ambient physical observables change under the deformation by defect operators.

In this effort, some of the most powerful tools that we have come from imposing symmetries on both the ambient and defect theories. The ambient field theories we consider are $6d$, supersymmetric, and invariant under $6d$ flat-space conformal symmetry $SO(6,2)$; superconformal field theories (SCFTs). The defects that we study in this work are supported on embedded co-dimension $2$ submanifolds, $\Sigma_4\hookrightarrow\CM_6$,  that will preserve at least $1/4$ of the total supersymmetries, i.e. $\CN\geq1$ $4d$ supersymmetry, as well as an $SO(4,2)\times U(1)_N\subset SO(6,2)$ global symmetry representing the defect conformal symmetry and $U(1)_N$ rotations in $\CM_6/\Sigma_4$.  We will refer to these theories as defect [super]conformal field theories (D[S]CFTs).

In the following, we will focus on ambient theories that are maximally superconformal $\CN=(2,0)$ SCFTs with gauge algebra $A_{N-1}$ in the large $N$ limit and the $1/4$- and $1/2$-BPS co-dimension $2$ defects that they support.  Despite the highly restrictive symmetries imposed, $6d$ $\CN=(2,0)$ SCFTs and their defect operators pose a challenge to direct study. We know from the worldvolume theory of a stack of coincident M5-branes \cite{Strominger:1995ac} or M5-branes probing $ADE$ singularities \cite{DelZotto:2014hpa} that $6d$ $\CN\geq(1,0)$ SCFTs exist, but generally they have no known Lagrangian description. We also know that $6d$ SCFTs constructed from M-theory support $4d$ BPS defect operators engineered at the intersection of orthogonal stacks of M5 branes. Since we often lack a Lagrangian description, our efforts to characterize these $\mf{d}=4$ defect (super)conformal field theories, or D(S)CFTs, are limited to analyzing their global properties using techniques such as anomaly inflow (e.g. \cite{Bah:2019jts}) and chiral algebra methods \cite{Beem:2014kka}.  That said, there is a tremendous amount that we can learn about the defect theory by studying its conformal anomalies.

As with any systems preserving an $SO(d,2)$ global conformal symmetry, putting the ambient theory on a curved $\CM_d$ results in a non-trivial Weyl anomaly.  Crucial to our understanding of DCFTs, the theory supported on $\Sigma_{\mf{d}}\hookrightarrow \CM_d$ has its own defect-localized contributions to the total Weyl anomaly that are sensitive to both the intrinsic submanifold geometry and its embedding in the ambient space.  The resulting defect Weyl anomaly can be far more complicated than that of an ordinary $\mf{d}$-dimensional theory. For example, it is common knowledge that the Weyl anomaly in $d=4$ is a combination of an `A-type' anomaly $\sim a E_4$, where $E_4$ is the $4d$ Euler density, and a `B-type' anomaly $\sim c |W|^2$ with $W_{\mu\nu\rho\sigma}$ denoting the Weyl-tensor \cite{Deser:1993yx}.  On the other hand, it was recently discovered in \cite{Chalabi:2021jud} that the Weyl anomaly of a $\mf{d}=4$ defect in an ambient theory with $d\geq6$ has a total of 29 terms\footnote{These 29 terms include 6 terms that break parity on the defect submanifold. The limit case of a co-dimension 1 defect in $5d$ has 12 (including 3 parity odd) terms in the Weyl anomaly \cite{FarajiAstaneh:2021foi,Chalabi:2021jud}.}. 

The challenge thus far has been finding tractable, non-trivial $\mf{d}=4$ defect systems beyond free theories (e.g. \cite{Bianchi:2021snj}) in which any of the 29 available defect Weyl anomalies can be computed\footnote{For probe branes wrapping AdS$_5\subset$ AdS$_{d+1}$, all of 23 of the parity even anomalies can be holographically computed \cite{Chalabi:2021jud} using the work of Graham and Reichert in \cite{Graham:2017bew}.}.  In light of recently discovered $11d$ supergravity (SUGRA) solutions that holographically describe certain $\mf{d}=4$ BPS defects in $6d$ SCFTs \cite{Gutperle:2022pgw,Gutperle:2023yrd}, we have a window on strongly coupled, non-Lagrangian defect systems that can be approached with standard tools in holography to compute quantities known to be controlled by defect anomalies.  

In this work, we study both the $1/4$-BPS `two-charge' solutions in $11d$ constructed as the uplift of domain wall solutions in $7d$ gauged SUGRA and the $1/2$-BPS `electrostatic' solutions for bubbling geometries \cite{Lin:2004nb, Lunin:2007ab} built along the lines of those in \cite{Bah:2021mzw,Bah:2022yjf} but with non-compact internal spaces. By holographically computing the one-point function of the stress-energy tensor and the flat defect contribution to the entanglement entropy (EE) of a spherical region co-original with the defect, we will be able to extract two of the 29 possible defect Weyl anomaly coefficients.  In doing so, we find two independent pieces of data that characterize these defect systems.

The final results of our analysis are collected in Tab.~\ref{tab:anomaly-summary}.  On the first line, the two-charge defect Weyl anomaly coefficients are expressed in terms of the charges $q_1,\,q_2$ and of the location $y_+$, itself expressible in terms of the charges, where the geometry of the SUGRA solution either smoothly caps off or ends on a conical deficit. On the second line, we display the anomalies for the electrostatic solutions, which holographically describe defects in the $6d$ $A_{N-1}$ $\CN=(2,0)$ SCFT labelled equivalently by a Lie algebra homomorphism $\vartheta:\mf{sl}(2)\to \mf{su}(N)$,  the choice of Levi subalgebra $\mf{l}\subset\mf{su}(N)$ associated with the Levi subgroup $L= S(U(N_1)\times\ldots U(N_n))$, or the Young diagram corresponding to the partition of $N = \sum_{a=1}^n N_a$. Additionally, the construction of the electrostatic solutions allows for $\mathbb{Z}_{k_a}$ orbifolds at the location of the $a^{{\rm th}}$ stack of M5-branes, which in dimensionally reducing to a $4d$ theory are interpreted as monopole charges for the $U(N_a)$ factors.  The parameters $N_a$ and $k_a$ completely determine both $a_\Sigma$ and $d_2$ for the electrostatic solutions.  

\bgroup
\def\arraystretch{1.35}
\begin{table}[t]
  \begin{tabular}{|c||c|c|}
    \hline
         & $a_\Sigma$ & $d_2$ \\\hline
      Two-charge  & $\frac{N^3}{24}(1-y_+^2)$ & $-\frac{N^3}{6}(q_1+q_2)$\\\hline
       Electrostatic  &$\frac{N^3}{32} -\frac{1}{96}\sum\limits_{a}\left(\frac{1+2k_a}{k_a^2}N_a^3 +\sum\limits_{b=a+1}^nN_ak_b\left(\frac{N_a^2}{k_a^2}+3\frac{N_b^2}{k_b^2}\right)\right)$ & $-\frac{1}{24}\left(N^3 -\sum\limits_a \frac{N_a^3}{ k_a^{2}}\right)$\\ \hline
    \end{tabular}
    \caption{Holographic defect Weyl anomaly coefficients for the 11d uplift of the $1/4$-BPS, two-charge domain wall solution and the $1/2$-BPS electrostatic solutions.}
    \label{tab:anomaly-summary}
\end{table}

The results that we obtain for these $\mf{d}=4$ defect anomalies have implications and open up questions beyond their roles in $6d$ SCFTs.   Under the (partially) twisted dimensional reduction on genus-$\bf{g}$ Riemann surface $\CC_{\bf{g}}$ either with $\Sigma_4$ wrapping two legs along $\CC_{\bf{g}}$ or with $\CC_{\bf{g}}$ orthogonal to $\Sigma_4$ in $\CM_6$, the theory containing a defect dual to the electrostatic solution descends to a $4d$ class$-\mc{S}$ $\CN=2$ SCFT \cite{Gaiotto:2009we,Alday:2009aq} deformed by a $\mf{d}=2$ surface defect \cite{Alday:2009fs,Gaiotto:2009fs} or by (possibly irregular \cite{Tachikawa:2009rb,Tachikawa:2010vg,Chacaltana:2012zy}) punctures on its UV curve.  For example in the case of $\Sigma_4$ wrapping $\CC_{\bf{g}} = \mathbb{T}^2$, the $4d$ description is of Gukov-Witten defects in $\CN=4$ $SU(N)$ super-Yang Mills theory \cite{Gukov:2006jk,Gukov:2008sn}, whose defect Weyl anomalies are known \cite{Gomis:2007fi,Drukker:2008wr,Jensen:2018rxu}.  While the precise map between the 29 defect anomalies and the two independent Weyl anomalies of a surface operator in $4d$ \cite{Bianchi:2019sxz} or the central charges of the $4d$ SCFT itself is unknown at present, our results provide some insight into how some of the defect data in $6d$ is reorganized into $4d$ (defect) Weyl anomalies.  For example, from Tab.~\ref{tab:anomaly-summary} and the results of \cite{Bah:2022yjf}, we see an interesting relation between the central charge $c_{\rm 4d}$ of the dimensionally reduced 4d $\CN=2$ SCFT and the A-type anomaly $1/2$-BPS co-dimension 2 defect in the $6d$ $A_{N-1}$ $\CN=(2,0)$ SCFT dual to the electrostatic solutions: $a_\Sigma = \frac{N^3}{32}+c_{\rm 4d}$.

Additionally, we evaluate the action of $11d$ SUGRA on the uplift of the two-charge solutions. The Hodge dual of the four-form flux is closed, which allows us to recast the on-shell action as an integral over the $10d$ boundary defined by $u=0$, where $u$ is the AdS$_7$ radial direction in Fefferman-Graham (FG) gauge. The volume of the $10d$ leaves normal to $u$ diverges as $u\to0$. In fact, it is divergent also at non-zero $u$, due to the infinite volume of the AdS$_5$ subspacetime contained within each leaf. We regulate the first divergence by subtracting from the on-shell action the contribution of the AdS$_7\times\mathds{S}^4$ vacuum, whose asymptotics match those of the uplift of the two-charge solutions. Finally, we regulate the polynomial divergences in the AdS$_5$ volume, and extract the prefactor of the remaining logarithmic divergence, which is a universal quantity. Our result is in given in \eq{Two-charge-on-shell-action}, which we reproduce here for the reader's convenience:
\begin{equation}
S_{\rm OS}^{\rm(ren)}\big|_{\log}=\frac{N^3(4q_1q_2 -2(q_1+q_2)y_+(1-y_+)+5y_+(1-y_+^2))}{1920y_+}~.
\end{equation}
At first sight, this result appears to be in tension with the computation of the on-shell action of $7d$ $\mathcal{N}=4$ gauged SUGRA performed in \cite{Gutperle:2022pgw}. The origin of this discrepancy can be identified in the parametrization of the bulk integrand. Indeed, due to the non-trivial nature of the fibration via which the $11d$ SUGRA solution is obtained, the requirement that the $11d$ spacetime be asymptotically locally in FG parametrization results in a redefinition of the angular coordinates which mixes the $U(1)$ isometry of the $7d$ spacetime and the $U(1)$ R-symmetry. In the $7d$ picture, this coordinate redefinition corresponds to a large gauge transformation which maps the theory to a singular gauge. Crucially, the uplift resolves this singularity, thus producing a solution which is perfectly regular from an $11d$ perspective. When performing the computation of the on-shell action in the original $7d$ coordinates, which do not manifest the asymptotic local AdS$_7\times\mathds{S}^4$ structure of the solution, \eq{Two-charge-on-shell-action} picks up a deformation which recovers precisely the result of \cite{Gutperle:2022pgw}, up to normalization factors. 

The present work is structured as follows. In \sn{Review}, we first review the pertinent aspects of Weyl anomalies for $4d$ defects and highlight their connection to physical quantities that we will compute in later sections.  We will also briefly review the solutions in $11d$ SUGRA that holographically describe $1/4$-BPS and $1/2$-BPS co-dimension 2 defects in $6d$ SCFTs.  In \sn{Holographic-Tij}, we compute the holographic stress-energy tensor one-point function for both the $1/4$-BPS, two-charge solution and a generic $1/2$-BPS electrostatic solution, which we use to find the defect B-type Weyl anomaly that we call $d_2$.  In \sn{Holographic-EE}, we holographically compute the defect contribution to the EE of a spherical region, which we use to determine defect A-type Weyl anomaly, $a_\Sigma$. In \sn{On-shell-action} we compute the on-shell action for the $11d$ uplift of the two-charge solutions and highlight a discrepancy with the same computation done in the domain wall description in $7d$ $\CN=4$ gauged SUGRA.  In \sn{Discussion}, we discuss comparisons to field theory results and future directions. 

In addition, we collect some useful intermediate results in the appendices. In \app{FG}, we detail the asymptotic maps of the metrics for the solutions we consider into Fefferman-Graham form.  Furthermore, in \app{On-shell-regulation}, we discuss the details of the regulating scheme for the on-shell action including the vacuum solution that we use in background subtraction as well as the renormalized volume of the AdS$_5$ geometry.

\section{Review}\label{sec:Review}

In this section, we will very briefly review some key background material in order to orient the subsequent computations. In the first subsection, we will introduce the two defect Weyl anomalies and discuss the physical quantities that they control, which will be the focus of the computations to follow.  In the second subsection, we will give a short overview of the two solutions to $11d$ SUGRA that will be the focus of our holographic study.

\subsection{Defect Weyl anomalies}\label{sec:Review-anomalies}

Up to a total derivative, the Weyl anomaly of an ordinary $4d$ CFT has two independent contributions\footnote{This basis is not unique, and one can exchange either $E_4$ or $W^2$ for Branson's $Q$-curvature \cite{Branson1991ExplicitFD} and a total derivative, which gives a basis for the $4d$ Weyl anomaly that is particularly convenient for holography.},
\begin{align}
    T^\mu{}_\mu = \frac{1}{4\pi^2}(-a_{4d}E_4 + c |W|^2).
\end{align}
The first term proportional to the Euler density $E_4$ is the so-called ``A-type'' anomaly in the classification of \cite{Deser:1993yx}, which exists in all even-dimensional CFTs and is unique in that it transforms as a total derivative under Weyl transformations.  The second term given by the square of the Weyl tensor is a ``B-type'' anomaly. In arbitrary even-dimensional CFTs, there is generally a tower of B-type anomalies each of which is exactly Weyl invariant and built out of non-topological, rank-$\frac{d}{2}$ monomials in curvatures. The Weyl anomaly coefficients of a 4$d$ CFT control correlation functions of the stress-energy tensor \cite{Osborn:1993cr}, and have strong upper and lower bounds on their ratio \cite{Hofman:2008ar}; $a_{4d}$ also appears in the EE \cite{Myers:2010tj}, and obeys an `$a$'-theorem under renormalization group (RG) flows \cite{Cardy:1988cwa,Komargodski:2011vj}. For 4$d$ SCFTs with an R-symmetry, $a_{4d}$ and $c$ are both related to the cubic and mixed R-anomalies through non-perturbative formulae \cite{Anselmi:1997am}.

The Weyl anomaly of a conformal defect supported on $\Sigma_{\mf{d}}\hookrightarrow\CM_d$ is much richer due to the additional freedom of building submanifold conformal invariants out of not only the intrinsic curvature but also the normal bundle curvature, the pullback of curvature tensors from the ambient space, and the second fundamental form for the embedding.  For conformal defects on $\Sigma_4\hookrightarrow \CM_d$ of co-dimension $2$ or greater\footnote{The limit case of co-dimension one is far more restricted and only leads to 9 parity even anomalies \cite{FarajiAstaneh:2021foi, Chalabi:2021jud}.}, there are a total of 23 anomalies respecting submanifold parity \cite{Chalabi:2021jud}\footnote{There are an additional 6 parity odd defect Weyl anomalies, but as of yet, there are neither any known physical quantities in which they appear nor any no-go theorem to forbid them. }.  The complete form of the 4$d$ defect Weyl anomaly is cumbersome, and so we will only display the parts relevant to the computations in the following sections (see eq.~3.1 of \cite{Chalabi:2021jud} for the full expression): 
\begin{align}
\begin{split}\label{eq:defect-Weyl-anomaly}
\hspace{-1cm}\left. T^\mu{}_{\mu}\right|_{\Sigma_4} \supset\frac{1}{(4\pi)^2}\Big(&-\bcc\overline{E}_4 
+d_{2}\CJ_2+\ldots
\Big)\,.
\end{split}\hspace{-2cm}
\end{align}
The first term is recognizable as the defect A-type anomaly proportional to the \textit{intrinsic} Euler density, $\overline{E}_4$, of $\Sigma_4$.  The second term $\CJ_2$ is a B-type anomaly built out of a complicated linear combination of the submanifold pullback of the ambient curvatures, connection on the normal bundle, normal bundle curvature, and the second fundamental form for the embedding (see eq.~3.2 of \cite{Chalabi:2021jud} for the full expression). Importantly, $\CJ_2$ does not contain a term like the pullback of $|W|^2$ or the square of the intrinsic Weyl tensor, and so is not analogous to the B-type anomaly of a standalone $4d$ CFT above.

While it is unclear what physics the vast majority of terms in the full expression of the defect Weyl anomaly control, the two anomalies displayed above appear in two physical quantities that will be the primary focus of the following work.

The first quantity we will analyze is the one-point function of the stress-energy tensor.  For a $\mf{d}$-dimensional conformal defect embedded in a $d$-dimensional CFT, conformal symmetry preserved by the defect constrains the form of the one-point function of the stress-energy tensor a distance $x_\perp$ away from the defect to be of the form
\cite{Kapustin:2005py, Billo:2016cpy}
\begin{align}\label{eq:stress-tensor-one-pt-fn}
\langle T^{ab} \rangle = - h_T \frac{(d-\mf{d}-1)\delta^{ab}}{|x_\perp|^d}\, ,\qquad \langle T^{ij} \rangle =h_T\frac{(\mf{d}+1)\delta^{ij} - d \frac{x^i_\perp x_\perp^j}{|x_\perp|^2}}{|x_\perp|^d}\,,
\end{align}
where $a,b$ index directions parallel to the defect and $i,j$ label directions normal to the defect. By starting from the defect geometry  $\Sigma_4=\mathbb{R}^4\hookrightarrow\mathbb{R}^d$ and then finding the totally transverse log divergent parts of the effective action in the presence of a linear ambient metric perturbation \cite{Lewkowycz:2014jia,Chalabi:2021jud}, it can be shown that  the normalization of the stress-energy tensor one-point function is determined by
\begin{equation}
\label{eq:h-d2-q}
 h_T = - \frac{ \, \Gamma \left(\frac{d}{2}-1\right)}{\pi ^{\frac{d}{2}}\,(d-1)} d_{2}\,.
\end{equation}
In the case that we are particularly interested in for the following work, i.e. $d=6$,  
\begin{equation}\label{eq:h-d2-q=2}
h_T = - \frac{1}{5 \pi^3} d_{2}\,.
\end{equation}
There is a constraint on the sign of $d_2$ that follows from the assumption that the average null energy condition (ANEC) holds in the presence of a defect. That is, the statement of the ANEC is that for any state $\left.|\Psi\right>$ of a QFT, the expectation value of the stress-energy tensor projected along a null direction $v^\mu$ in that state satisfies
\begin{align}\label{eq:ANEC}
    \int_{-\infty}^\infty d\lambda\ \left<\Psi|T_{\mu\nu}|\Psi\right>v^\mu v^\nu \geq 0~,
\end{align}
where $\lambda$ parametrizes the null geodesic. From \eq{h-d2-q}, we see that by taking the ambient theory to be a CFT and $\left.|\Psi\right>$ to be the vacuum state of the theory deformed by a defect and orienting the null ray $v^\mu$ to be parallel to the defect and separated by a distance $x_\perp$ in the normal direction, $h\geq 0$, which implies $d_2\leq 0$ \cite{Jensen:2018rxu,Chalabi:2021jud}\footnote{ In fact, it has recently been argued that the quantum null energy condition (QNEC), which is a stronger energy condition valid in any ambient QFT and reduces to ANEC in a certain limit (see e.g. \cite{Faulkner:2018zfg}), holds in the presence of a defect \cite{Casini:2023kyj}; putting $d_2\leq 0$ and any other sign constraint derived from such energy conditions on even firmer ground.}.  

The other physical quantity controlled by defect Weyl anomalies that we will study below is the contribution to the EE of a spherical region of size $R$ centered on $\Sigma_4=\mathbb{R}^{1,3}\hookrightarrow\mathbb{R}^{1,d-1}$.  Following the same logic that formed the basis of the proof for $2d$ defects \cite{Kobayashi:2018lil,Jensen:2018rxu}, it was shown in \cite{Chalabi:2021jud} that for a $4d$ defect of co-dimension $d-4$, the coefficient of the universal, i.e. the log divergent, part of the defect EE is 
\begin{equation}\label{eq:defect-EE}
S_{\rm EE}[\Sigma]\Big|_{\log}= -4\left[ \, a_{\Sigma} + \frac{1}{4}\frac{ (d-4)(d-5)}{d-1} \, d_2 \right]\log \left(\frac{R}{\e}\right),
\end{equation}
where $\e\ll R$ is a UV cutoff scale and $\big|_{\log}$ denotes dropping the leading non-universal divergences as well as the trailing scheme-dependent terms.

For a conformal defect on $\Sigma_4$, we will use a background subtraction scheme to isolate the defect contribution to the EE.  That is, our computations below will use
\begin{align}\label{eq:defect-EE-anomalies}
    4a_\Sigma +\frac{2}{5}d_2 = -R\pd_R\left(S_{\rm EE}[\Sigma] - S_{\rm EE}[\emptyset]\right)|_{R\to 0}~,
\end{align}
where $S_{\rm EE}[\emptyset]$ is the EE computed without the defect, i.e. the EE of the vacuum of the $6d$ ambient theory. Thus, combining the computation of $d_2$ from $\Delta\left<T_{ij}\right>$ with the result of \eq{defect-EE-anomalies}, we can compute the defect A-type anomaly unambiguously.

Unlike $d_2$, however, there is no constraint on the sign of $a_\Sigma$.  Indeed, in the simple case of a free scalar on a $5d$ manifold with a boundary, $a_\Sigma>0$ for Neumann (Robin) boundary conditions, while $a_\Sigma<0$ for Dirichlet \cite{FarajiAstaneh:2021foi}\footnote{Note we are using the conventions for the definition of the $4d$ defect A-type anomaly $a_{\Sigma}$ as in \cite{Chalabi:2021jud}, which differs from the defect A-type anomaly, $a$, in \cite{FarajiAstaneh:2021foi} by $a_\Sigma \leftrightarrow -a/5760$.}.  

\subsection{11d SUGRA solutions}\label{sec:Review-SUGRA}
\subsubsection*{Two-charge solutions}

We now briefly review the domain wall solutions in $7d$ $\CN=4$ gauged SUGRA found in \cite{Gutperle:2022pgw} and uplifted to $11d$ in \cite{Gutperle:2023yrd}. The bosonic $7d$ gauged SUGRA action built from the metric $g$, two scalars $\Phi_{1,2}$ and two $U(1)$ gauge fields $A_{1,2}$ takes the following form:
\begin{align}\label{eq:7d-gauged-SUGRA-action}
    S = -\frac{1}{16\pi G_{N}^{(7)}}\int d^7x\sqrt{|g|}\left(\CR  -\frac{1}{2}|\pd_\mu\Phi_I|^2 - \hat{g}^2V(\Phi) - \frac{1}{4} \sum_{I=1}^2 e^{\vec{a}_I \vec{\Phi}}F_I^2 \right).
\end{align}
Using $\vec{a}_1 = (\sqrt{2},\sqrt{2/5})$, $\vec{a}_2 =(-\sqrt{2}, \sqrt{2/5})$, the potential is given by
\begin{align}
    V = - 4 e^{-\frac{1}{2}(\vec{a}_1+\vec{a}_2)\vec{\Phi}}-2\left(e^{\frac{1}{2} (\vec{a}_1 +2\vec{\a}_2)\vec{\Phi}} +e^{\frac{1}{2} (2\vec{a}_1 +\vec{\a}_2)\vec{\Phi}}\right) +\frac{1}{2}e^{2(\vec{a}_1+\vec{a}_2)\vec{\Phi}}~.
\end{align}
The domain wall solution to \eq{7d-gauged-SUGRA-action} describing the double analytic continuation of a charged black hole is given by
\begin{align}
    ds_7^2 = (yP(y))^{\frac{1}{5}} ds_{\rm AdS_5}^2+\frac{y(yP(y))^{\frac{1}{5}}}{4Q(y)}dy^2+\frac{yQ(y)}{(yP(y))^{\frac{4}{5}}}dz^2,
\end{align}
where the polynomials $P,\,Q$ are given by
\begin{subequations}
    \begin{align}\label{eq:Two-charge-P}
P(y)& = H_1(y)H_2(y),\\\label{eq:Two-charge-Q}
Q(y) &= -y^3 +\mu y^2 +\frac{\hat{g}^2}{4}P(y),
\end{align}
\end{subequations}
where $ H_I(y) = y^2+q_I$, $I\in\{1,2\}$.
The gauge fields in this solution\footnote{Note that, in general, the action in \eq{7d-gauged-SUGRA-action} does not qualify as a consistent truncation of $11d$ supergravity. The $7d$ solutions considered here, however, are characterized by $F_1\wedge F_2=0$; this guarantees that their uplift produces consistent solutions of the $11d$ theory \cite{Cvetic:1999xp}. 
} are given by
\begin{align}
    A_I = \left(\sqrt{1-\frac{\mu}{q_I}}\frac{q_I}{H_I(y)} +a_I\right)dz~.
\end{align}
In order to find BPS solutions, SUSY forces $\mu=0$.  For both $q_I\neq 0$, the solutions are $1/4$-BPS, while setting one charge, say $q_2$, to zero allows for $1/2$-BPS solutions.  In the following, we will refer to the former $1/4$-BPS cases as `two-charge solutions' and the latter $1/2$-BPS cases as `one-charge solutions'.  The coordinate \(y\) ranges from  \(y_+\), the largest root of \(Q(y)\), to infinity.  To have a smooth geometry one can choose the gauge so that $A_I(y_+)=0$ by appropriate choice of the \(a_I\). Setting $\hat{g}=2$, the AdS$_5\times\mathds{S}^1$ geometry does not have a conical deficit provided $z\in[0,2\pi)$ (this will be assumed in the uplift to $11d$).  At \(y=y_+\), the geometry either has a smooth cap or a conical deficit \(2\pi\frac{\hat{n}-1}{\hat{n}}\) with \(\hat n\) related to \(y_+\) by the constraint \(\hat n \, Q'(y_+) = y_+^2\).

The conditions \(Q(y_+)=0\) and \(\hat n \, Q'(y_+) = y_+^2\) can be solved to determine \(q_1\) and \(q_2\) in terms of \(\hat n\) and \(y_+\) as follows
\begin{align}\label{eq:Two-charge-qI}
    q_I = y_+ \left( \frac{3 \hat n+ 1}{\hat n\hat{g}^2} - y_+
    \pm \frac{2}{\hat{g}} \sqrt{\frac{(1+3 \hat n)^2}{4 \hat{g}^2 \hat{n}^2} - y_+} \right),
\end{align}
where \(q_1\) and \(q_2\) are chosen with opposite signs for the square root.  This has real solutions provided \(0 \leq y_+ \leq y_{+,\text{max}}\) with \( y_{+,\text{max}} = (1+3 \hat n)^2/4 \hat{g}^2 \hat{n}^2\).  It will be useful later to notice that the sum \(q_1 + q_2\) is always non-negative as is evident from
\begin{align} \label{eq:q1q2positivity}
    \frac{q_1 + q_2}{2 y_+} = \left( \frac{3\hat n + 1}{\hat n\hat{g}^2} - y_+ \right) \geq \left( \frac{3\hat n + 1}{\hat n\hat{g}^2} - y_{+,\text{max}} \right)
    = \frac{(\hat n - 1)(3 \hat n+1)}{4 \hat{g}^2 \hat n^2} \geq 0.
\end{align}

The $7d$ solutions above were uplifted in \cite{Gutperle:2023yrd} using the standard ansatz for the Kaluza-Klein (KK) reduction of $11d$ SUGRA on $\mathds{S}^4$ \cite{Cvetic:1999xp}; namely, the $11d$ metric and four-form flux are given by the $7d$ data as follows,
\begin{subequations}\label{eq:Two-charge-uplift}
\begin{align}
ds^2_{11}&=\tilde{\Delta}^{1/3}ds^2_7+\hat{g}^{-2}\tilde\Delta^{-2/3}\left[X_0^{-1}d\mu_0^2+\sum_{I=1}^2X_I^{-1}\left(d\mu_I^2+\mu_I^2(d\phi_I+\hat{g}A_I)^2\right)\right]\\
\star_{11}F_4&=2\hat{g}\sum_{i=0}^2\left(X_i^2\mu_i^2-\tilde\Delta X_i\right)\Upsilon_7+\hat{g}\tilde\Delta X_0\Upsilon_7+\frac{1}{2\hat{g}}\sum_{i=0}^2\star_7d\ln X_i\wedge d(\mu_i^2)\\
&\quad+\frac{1}{2\hat{g}^2}\sum_{I=1}^2X_I^{-2}d(\mu_I)^2\wedge(d\phi_I+\hat{g} A_I)\wedge\star_7 F_I \nonumber
\end{align}
\end{subequations}
where $\Upsilon_7$ is the $7d$ volume form and we defined
\begin{align}
   &X_1 
   = \frac{(yH_2(y))^{\frac{2}{5}}}{H_1(y)^{\frac{3}{5}}}~,&
   &X_2 
   = \frac{(yH_1(y))^{\frac{2}{5}}}{H_2(y)^{\frac{3}{5}}}~,&
   &X_0 = (X_1X_2)^{-2}~,&
   &\tilde\Delta=\sum_{i=1}^2X_i\mu_i^2~,&
\end{align}
as well as
\begin{align}
    &\mu_0=\sin\psi\cos\zeta~,&
    &\mu_1=\sin\zeta~,&
    &\mu_2=\cos\psi\cos\zeta~.&
\end{align}

Explicitly, the $11d$ metric above for the two-charge $1/4$-BPS solutions can be brought into the form
\begin{align}\label{eq:Two-charge-metric}
\begin{split}
      ds_{11}^2 &= \hat{f}^2_{\rm AdS}ds_{\rm AdS_5}^2 + \hat{f}_y^2dy^2 +\hat{f}_z^2dz^2 +\hat{f}_{\phi_i}^2d\phi_i^2 + \hat{f}_{z\phi_i}^2dz d\phi_i +\hat{f}_\psi^2d\psi^2 + \hat{f}_\zeta^2d\zeta^2 +\hat{f}_{\psi\zeta}d\psi d\zeta~,
\end{split}
\end{align}
where each of the $\hat{f}$'s displayed in \eq{Two-charge-fs} is a function of the $y,\,\psi,$ and $\zeta$ coordinates and also depends on the $q_I$'s and $a_I$'s. Note that in \eq{Two-charge-fs}, we have introduced the slightly abusive shorthand
\begin{align}
    \sin x \equiv s_x~,\qquad \cos x\equiv c_x~,
\end{align}
in order to compactly express some of the more cumbersome expressions, and we will adopt this notation throughout the following sections. Continuing on, the uplifted four-form field strength can be inferred from  
\begin{align}
\hspace{-.35cm}\frac{\star_{11}F_4}{\k^2} =&-2(\hat{H}(X_0+2(X_1+X_2))-2X_0^2+2(X_0^2-X_1^2)s_\zeta^2+2(X_0^2-X_2^2)c_\psi^2c_\zeta^2)\Upsilon_7\label{eq:Two-charge-potential}\\\nonumber
&+\frac{c_\zeta^2c_\psi s_\psi}{2X_0X_2}(X_2 \star_7dX_0-X_0\star_7dX_2)\wedge d\psi +\frac{c_\zeta s_\zeta}{2X_1}\star_7 dX_1\wedge d\zeta \\\nonumber
&- \frac{c_\zeta s_\zeta}{2X_0X_2}(X_2s_\psi^2 \star_7 dX_0 + X_0 c_\psi^2 \star_7 dX_2)\wedge d\zeta+\frac{c_\zeta s_\zeta}{4X_1^2}d\zeta\wedge(d\phi_1+2A_1)\wedge\star_7dA_1\\
&-\frac{c_\zeta c_\psi}{4X_2^2}(c_\zeta s_\psi d\psi +s_\zeta c_\psi d\zeta)\wedge(d\phi_2+2A_2) \wedge\star_7 dA_2\nonumber
\end{align}
where we have set $\hat{g}=2$. We also defined
\begin{align}
    \hat{H} = \frac{X_2(H_2-q_2c_\psi^2)c_\zeta^2}{y^2}+X_1s_\zeta^2~.
\end{align}

\subsubsection*{Electrostatic solutions}

In this subsection, we review the construction of an infinite class of `bubbling' solutions to $11d$ SUGRA with AdS$_5\times \mathds{S}^1$ boundary geometries that holographically describe $1/2$-BPS co-dimension 2 defects in $6d$  SCFTs \cite{Gutperle:2023yrd}.  There is a long history of AdS$_5$ compactifications in $11d$ SUGRA and M-theory holographically dual to $4d$ $\CN=2$ SCFTs, e.g. \cite{Maldacena:2000mw,Lin:2004nb,Lunin:2007ab,Gaiotto:2009gz}. The class into which the solutions of \cite{Gutperle:2022pgw, Gutperle:2023yrd} are embedded are a particular type of Lin-Lunin-Maldacena (LLM) `bubbling' geometries \cite{Lin:2004nb,Lunin:2007ab}.  

Recall that the general LLM solution consists of an $11d$ geometry with a warped product AdS$_5\times\mathds{S}^2$ over $\CM_4$ realized as a $U(1)_\chi$-fibration over a $3d$ base space $\CB_3$ supported by four-form flux. The data that specifies the solution is encoded in a function that satisfies a non-linear Toda equation on $\CB_3$, which is generically difficult to solve.  However, by imposing that $\CB_3$ has an additional $U(1)_\beta$ isometry, the Toda equation can be cast in an axi-symmetric form that can be solved more easily.  Further facilitating finding general solutions to the axi-symmetric Toda equation, one can perform a B\"acklund transformation to map to a Laplace-type equation on $\mathbb{R}^3$, and so the problem is turned into an `electrostatic' one \cite{Ward:1990qt,Donos:2010va,Petropoulos:2014rva,Bah:2019jts,Bah:2022yjf,Gutperle:2023yrd}. Hence, the class of bubbling geometries reviewed below will be referred to as `electrostatic solutions' in the following sections.

In the formulation as a Laplace-type equation, finding a solution to the SUGRA equations of motion amounts to specifying a linear charge density $\varpi$ which determines the electrostatic potential $V$. Exploiting the axial symmetry of the problem on $\CB_3$, we take $\varpi=\varpi(\eta)$ to be aligned along the $\eta$-axis, i.e. the fixed point of the $U(1)_\beta$ rotations. The bosonic sector of these solutions takes the form
\begin{subequations}
    \begin{align}\label{eq:LLM-defect-metric}
    ds_{11}^2 &= \kappa_{11}^{\frac{2}{3}}\left(\frac{\dot{V} \sigma}{2V^{\prime\prime}}\right)^{\frac{1}{3}}\Bigg(4ds_{\rm AdS_5}^2 + \frac{2V^{\prime\prime}\dot{V}}{\sigma}d\Omega_2^2 + \frac{2(2\dot{V}-\ddot{V})}{\dot{V}\sigma}\Big(d\beta + \frac{2\dot{V}\dot{V}^\prime}{2\dot{V}-\ddot{V}}d\chi\Big)^2\\\nonumber
    &\hspace{2.75cm}+\frac{2V^{\prime\prime}}{\dot{V}}\Big(dr^2 +\frac{2\dot{V}}{2\dot{V}-\ddot{V}}r^2d\chi^2 + d\eta^2\Big)\Bigg)\\\nonumber
    &\equiv f_{\rm AdS}^2ds_{\rm AdS_5}^2 + f_{\mathds{S}^2}d\Omega_2^2 + f_\beta^2d\beta^2 +f_\chi^2d\chi^2 + f_{\beta\chi}^2d\beta d\chi + f_3^2(dr^2+d\eta^2)~,\\
    \label{eq:LLM-defect-C3}
    C_3 &= \frac{2\kappa_{11}}{\sigma}\left(\left(\dot{V}\dot{V}^\prime -\sigma\eta\right)d\beta -2\dot{V}^2V^{\prime\prime}d\chi\right)\wedge \Upsilon_{\mathds{S}^2}~,
\end{align}
\end{subequations}
where we have adopted the notation where ${\Upsilon_{\CM}:= \sqrt{|g_{\CM}|}dx^1\wedge\ldots\wedge dx^d}$ is the volume form on a $d$-dimensional manifold $\CM$.  In this notation, the coordinates $\{r,\eta,\beta\}$ span $\CB_3$, $\kappa_{11} = \pi \ell_P^3/2$, and 
\begin{align}
    V^\prime \equiv \pd_\eta V,\qquad \dot{V} \equiv r\pd_r(V),\qquad \sigma\equiv V^{\prime\prime}(2\dot{V}-\ddot{V})+(\dot{V}^\prime)^2.
\end{align}

In this background, away from sources, the electrostatic potential $V(r,\eta)$ satisfies
\begin{equation}\label{eq:Electrostatic-Laplace}
    \ddot{V}(r,\eta)+r^2V^{\prime\prime}(r,\eta)=0,
\end{equation}
subject to the boundary condition $\partial_rV|_{\eta=0}=0$. Exploiting the $U(1)_\beta$ isometry imposed on $\CB_3$, the line charge distribution $\varpi(\eta)$ specifying the solution is related to the Laplace potential $V$ by
\begin{equation}\label{eq:varpi}
    \varpi(\eta)=\lim_{r\to0^+}\dot{V}(r,\eta).
\end{equation}
Given an appropriate $\varpi(\eta)$, the solution to \eq{Electrostatic-Laplace} can be expressed in terms of a Green's function, $G(r,\eta,\eta^\prime)$, as
\begin{align}
    V(r,\eta) = -\frac{1}{2}\int d\eta^\prime G(r,\eta,\eta^\prime)\varpi(\eta^\prime).
\end{align}
By the symmetry of the problem, the Green's function can be written simply using the method of images as \cite{Bah:2022yjf,Gutperle:2023yrd}
\begin{align}
    G(r,\eta,\eta^\prime) = \frac{1}{\sqrt{r^2+(\eta-\eta^\prime)^2}}-\frac{1}{\sqrt{r^2+(\eta+\eta^\prime)^2}}.
\end{align}
The complete description of the solution to the $11d$ SUGRA field equations is thus given by finding a $\varpi(\eta)$ that obeys a set of necessary conditions.

For a generic $\varpi(\eta)$, the constraints that follow from charge conservation and regularity (modulo $A_k$ singularities on $\CM_4$) of the full $11d$ geometry were given in \cite{Gaiotto:2009gz}.  Satisfying these constraints determines the profile of $\varpi(\eta)$ to be a continuous, convex piecewise linear function of $\eta$ with integer slope, whose slope decreases by integer values at discrete $\eta_a$.  In general, the boundary conditions and symmetry imposed on $V$ in solving \eq{Electrostatic-Laplace} require $\varpi(0)=0$.  However, there are generally two cases for the behavior of $\varpi$ as $\eta$ increases.

In the first case, apart from the zero at the origin, $\varpi$ has a zero at some value $\eta=\eta_c>0$ where the internal space closes off.  The geometry of the $11d$ SUGRA solution is then a warped product of AdS$_5$ over the compact internal space $\CM_6 = \mc{C}_{\bf{g}}\times \CM_4$, and holographically describes a $4d$ theory that descends from the compactification of a $6d$ SCFT on a Riemann surface $\mc{C}_{\bf{g}}$.  The generic charge distribution is decomposed into $n+1$ `regular' intervals with positive slope and an `irregular' interval $[\eta_n,\eta_c]$ with negative slope fixed by ratios of four-form flux.  The data associated with the kinks between the regular parts of the charge distribution, namely a partition of $N$, label a regular puncture on $\CC_{\bf{g}}$, while the data specifying the slope of the irregular interval is mapped to an irregular puncture \cite{Bah:2022yjf}. This construction -- reminiscent of  other spindle compactifications engineering $4d$ SCFTs \cite{Ferrero:2020laf,Ferrero:2020twa, Boido:2021szx,Bah:2021mzw,Ferrero:2021wvk,Suh:2021ifj,Ferrero:2021etw,Giri:2021xta,Suh:2022olh} 
-- was argued in \cite{Bah:2022yjf} to be the SUGRA dual to class-$\CS$ constructions \cite{Gaiotto:2009we} of certain classes of Argyres-Douglas theories \cite{Argyres:1995jj} by analyzing anomalies and counting of Coulomb and Higgs branch operators in the field theory. While we will not study these types of solutions further here, we will mention some of their properties as they pertain to the results of holographic calculations of defect anomalies.

The second case, relevant for our study, is where $\varpi(\eta)$ has non-trivial support over the whole range $\eta\in[0,\infty)$ \cite{Gutperle:2023yrd}.  Since $\varpi(\eta)$ never turns around to hit the $\eta$-axis, the geometry $\CM_6$ in the $11d$ SUGRA solution is non-compact, and the $11d$ geometry can be engineered to be asymptotically locally AdS$_7\times\mathds{S}^4$ where the geometry of the conformal boundary of the AdS$_7$ factor is AdS$_5\times \mathds{S}^1$.  These solutions are, thus, interpreted as holographically describing co-dimension $2$ defect operators in 6$d$ SCFTs, where the defect operator `lives' at the conformal boundary of AdS$_5$.

As a simple example of a line charge density that gives rise to a non-compact geometry, it was shown in \cite{Gutperle:2023yrd} that the one-charge solution reviewed in the previous subsection can be recast in the language of the electrostatic solutions as a $\varpi(\eta)$ with two segments:
\begin{equation}\label{eq:singlekink}
    \varpi(\eta)=\begin{cases}
    \left(1+\frac{1}{\sqrt{1-4q_1}}\right)\eta, & \quad \eta\in\big[0,\frac{N}{2}\sqrt{1-4q_1}\big] \\ \eta+N/2, & \quad \eta\in\big[\frac{N}{2}\sqrt{1-4q_1} ,\infty\big).
    \end{cases}
\end{equation}
Due to $\varpi$ being continuous and piecewise linear, we will refer to the solution engineered by \eq{singlekink} as a `single kink solution'. This relation between the $q_2\to 0$ limit of the two-charge solutions and the simple single kink line charge distribution for the electrostatic solutions will be useful in later sections as a consistency check for our computations. We should also note that the constraint that the change in slope of $\varpi(\eta)$ is integral forces $q_1=\frac{j^2-1}{4j^2} $ for $j\in\mathbb{N}$.

Generalizing beyond the single kink solutions, the constraints on $\varpi(\eta)$ realizing a defect solution allow for a generic $n$-kink charge profile. Since $\varpi(\eta)$ is piecewise linear, its behavior on the $a^{\text{th}}$ interval, where $\eta\in[\eta_a,\eta_{a+1}]$ and $a\in\{0,1,\dots,n\}$, can be written as \cite{Bah:2019jts, Gutperle:2023yrd}
\begin{align}\label{eq:n-kink}
\varpi_a(\eta)&=\left(1+\sum_{b=a+1}^nk_b\right)\eta+\sum_{b=1}^a\eta_bk_b\\\nonumber
&\equiv p_{a+1}\eta + \delta_{a+1},
\end{align}
where in the second line we have introduced a convenient short hand for the slope $p_{a+1}$ and intercept $\delta_{a+1}$ of the line continued from the $a^{\text{th}}$ segment. From the boundary condition $\varpi(0)=0$ it is understood that $\eta_0=0$, and due to the semi-infinite domain of support we take $\eta_{n+1}\to\infty$. 

As a simple visualization of an arbitrary distribution, see the left side of \fig{chargedistribution}. Note that from the constraint following from the quantization of four-form flux $N = 2\sum_{a=1}^n\eta_ak_a$ along with the quantization of the $\eta_a$ and their ordering along the $\eta$-axis {($0<\ldots<\eta_{a}<\eta_{a+1}<\ldots< \eta_n$)}, there is a natural interpretation of the data $(\eta_{a}, k_a)$ specifying the charge distribution as a Young diagram, which is displayed on the right side of \fig{chargedistribution}. 

In the language of the field theory description, the Young diagram corresponding to the specific $\varpi(\eta)$ is in correspondence to both the Lie algebra homomorphism $\vartheta:\mathfrak{sl}(2)\to\mathfrak{g}$ and to the choice of Levi subalgebra $\mathfrak{l}$ of the $A_{N-1}$ gauge algebra. Furthermore, the slope change $k_a\in\mathbb{Z}$ between the $(a-1)^{\text{th}}$ and $a^{\text{th}}$ intervals corresponds to the monopole charge at the $\mathbb{R}^4/\mathbb{Z}_{k_a}$ orbifold point located at $(r,\eta)=(0,\eta_a)$ in the internal manifold. These points are the holographic realization of the non-Abelian summands $\mathfrak{su}(k_a)$ of the global symmetry algebra \cite{Gaiotto:2009gz}.

\begin{figure}[t]
    \centering
    \includegraphics[width=1.025\textwidth]{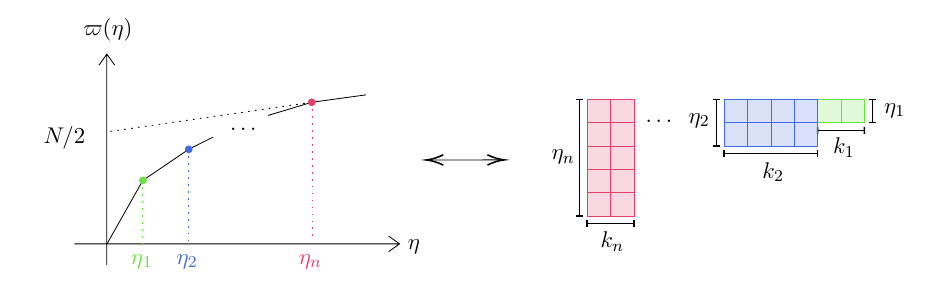}
    \caption{
 \textbf{(Left)} A generic line charge distribution $\varpi(\eta)$, with $n$ kinks at positions $\eta_a$ along the axis of cylindrical symmetry, specifying a solution to the axially symmetric Laplace equation in $\mathbb{R}^3$. The AdS$_7\times\mathds{S}^4$ vacuum corresponds to the single-kink ($n=1$) charge distribution with $k_1=1$; the location of the kink is then given by $\eta_1=N/2$. \textbf{(Right)} The Young Tableau corresponding to the partition $N=2\sum_{a=1}^nk_a\eta_a = \sum_{a=1}^nN_a$. The height and width of the $a$-th block are given by the location $\eta_a\in\mathbb{Z}$ and slope change $k_a\in\mathbb{Z}$ of the $a$-th kink in $\varpi(\eta)$, respectively. The AdS$_7\times\mathds{S}^4$ vacuum is associated to the $\mathbf{1}$ of $\mf{su}(N)$ determined by $n=k_1=1$ and $\eta_1=N/2$. 
    }
    \label{fig:chargedistribution}
\end{figure}

Lastly, for use in future computations, it will be convenient to define the `moments' of the potential  as in \cite{Gutperle:2023yrd} 
\begin{align}\label{eq:V-moments}
    m_j = \sum_{a=1}^n (p_a -p_{a+1})\eta_a^j =  \sum_{a=1}^n k_a\eta_a^j.
\end{align}
For most of the following, we will only need the first and third moments
\begin{align}
m_1 = \frac{N}{2}\qquad\text{and}\qquad m_3 = \sum_{a} \frac{N_a^3}{8k_a^2}
\end{align}
 respectively.
\section{Holographic stress-energy tensor one-point function} \label{sec:Holographic-Tij}

In this section, we will compute the contribution of a co-dimension 2 defect to the one-point function of the stress-energy tensor of the ambient $6d$ SCFT. To do so, we will reduce the $11d$ SUGRA backgrounds described in the previous section on the internal $\mathds{S}^4$ and employ the holographic renormalization methods of \cite{deHaro:2000vlm}. In their original formulation, these methods are meant to apply to asymptotically AdS solutions of pure Einstein-Hilbert gravity; therefore, we must ensure that the presence of the four-form flux in the dimensionally reduced M-theory solutions does not necessitate a modification of those methods. In both two-charge and electrostatic solutions, we will show that the field strength decays sufficiently fast as the conformal boundary of AdS$_7$ is approached so that it produces a vanishing contribution to the field equations on the boundary.

Following the general procedure in \cite{deHaro:2000vlm}, we begin by recasting the $11d$ metric as a perturbation $h_{11}$ about the AdS$_7\times\mathds{S}^4$ vacuum:
\begin{equation}
    ds^2_{11}=g_{{\rm AdS}_7\times\mathds{S}^4}+h_{11}.
\end{equation}
Dimensionally reducing on the internal $\mathds{S}^4$ then leads to the $7d$ line element
\begin{equation}\label{eq:KK-reduced-metric}
    ds^2_{7}=\left(1+\frac{\bar\varsigma}{5}\right)g_{\rm AdS_7}+\bar{h}_7,
\end{equation}
where $g_{\rm AdS_7}$ is the metric on AdS$_7$, the $7d$ field $h_7$ captures the fluctuations about the AdS$_7$ geometry, and $\varsigma$ is the trace of the fluctuations in the internal manifold. Bars indicate zero modes on the internal space; for instance\footnote{Our index conventions in this section are that $\mu,\nu,\dots$ are AdS$_7$ indices, $a,b,\dots$ are $\mathds{S}^4$ indices, and $i,j,\dots$ are $6d$ indices on the conformal boundary of AdS$_7$.},
\begin{align}
    \bar{\varsigma} = \frac{3}{4}\int_{\mathds{S}^4} \sqrt{g_{\mathds{S}^4}}\ h^{ab}g^{(0)}_{ab}.
\end{align}
Mapping the $7d$ line element into Fefferman-Graham (FG) gauge,
\begin{equation}
    ds^2_{7}=\frac{L^2}{u^2}\left(du^2+g\right),
\end{equation}
where the $6d$ metric $g$ admits the power series expansion
\begin{equation}
    g=g_{(0)}+g_{(2)}u^2+g_{(4)}u^4+g_{(6)}u^6+h_{(6)}u^6\log u^2+\dots,
\end{equation}
the $6d$ stress-energy tensor one-point function can be computed 
\begin{subequations}
    \begin{align}\label{eq:Tij-one-point-function}
    \langle T_{ij}\rangle\ dx^idx^j &= \frac{3L^5}{8\pi G_N^{(7)}}\left(g_{(6)} - A_{(6)} +\frac{S}{24}\right) \\
    &=\frac{N^3}{4\pi^3}\left(g_{(6)} - A_{(6)} +\frac{S}{24}\right),
\end{align}
\end{subequations}
where $A_{(6)}$ and $S$ are rank-2 tensors built out of $g_{(0)}$, and in the second line we have used the holographic map to field theory quantities 
\begin{align}
    \frac{1}{G_N^{(7)}} = \frac{\vol{\mathds{S}^4}}{G_{N}^{(11)}},\quad G_{N}^{(11)} = 2^4\pi^7\ell_P^9,\quad L^3= \pi N \ell_P^3, \quad \vol{\mathds{S}^4} = \frac{L^4\pi^2}{6}.
\end{align}
Note that in our conventions the internal $\mathds{S}^4$ has curvature scale $L^2/4$. Explicit expressions for $A_{(6)}$ and $S$ are provided in \cite{deHaro:2000vlm}\footnote{Note that the differences in sign are due the fact that we are using the convention that, in units of $L^{2d}$, the scalar curvature $\CR<0$ for a space of constant ``negative curvature''; whereas the authors of \cite{deHaro:2000vlm} use the opposite convention, $\CR>0$.}. Once the appropriate vacuum subtraction is performed, the defect contribution to $h_T$, and therefore to $d_2$, can be extracted via \eq{stress-tensor-one-pt-fn} and \eq{h-d2-q=2}.

\subsection{Two-charge solutions}\label{sec:Holographic-Tij-Two-charge}
In this subsection, we will focus on the $11d$ uplift of the two-charge solutions described in \ref{sec:Review-SUGRA} and compute $\left<T_{ij}\right>$ with the methods described above. In order to isolate the contributions from the holographic dual to the defect, we will employ a background subtraction scheme where we remove the contributions from vacuum AdS$_7\times \mathds{S}^4$. 

Before jumping in to the computation of $\left<T_{ij}\right>$, we need to carefully check that we can properly utilize our chosen holographic renormalization scheme. One of the crucial assumptions in the construction of \eq{Tij-one-point-function} is that Einstein's equations near the boundary of the dimensionally reduced AdS$_7$ geometry are not modified by contributions coming from non-trivial fluxes, such as the four-form curvature $F_4$. So, we must be careful to make sure that in the asymptotic small $u$ region, the components of the variation of the $F_{MNPQ}F^{MNPQ}$ part of the $11d$ SUGRA action involving AdS$_7$ directions fall off sufficiently fast so as to not modify the boundary equations of motion.  

 For the solutions in eqs.~(\ref{eq:Two-charge-metric}) and (\ref{eq:Two-charge-potential}), it suffices to show the fall-off conditions for the single charge case. Setting $q_2\to0$ and $a_2\to0$, transforming $\phi_I\to\varphi_I-2a_Iz$, and mapping to FG gauge as in \app{Two-charge-FG}, a quick computation shows the small $u$ behavior to be (up to overall numerical prefactors) 
\begin{align}\allowdisplaybreaks
    \begin{split}
        F_{a}^{MNP}F_{b MNP}&\sim c_\theta^2 g_{ab}+\ldots~,\\
        F_{\varphi_1}^{MNP}F_{\varphi_1 MNP}&\sim s_\theta^2+\ldots~, \\
        F_{\theta}^{MNP}F_{\theta MNP} &\sim1 +\ldots~, \\
        F_{z}^{MNP}F_{zMNP} &\sim q_1^2(13-5c_{2\theta})u^8+\ldots,
        \\F_{z}^{MNP}F_{\varphi_1 MNP} &\sim q_1 s_\theta^2 u^4+\ldots~,
        \\
        F_{y}^{MNP}F_{yMNP} &\sim q_1^2s_{2\theta}^2u^{12} +\ldots~,
    \end{split}
\end{align}
 where $g_{ab}$ are components along the $\mathds{S}^2\subset \mathds{S}^4$ and the AdS$_5$ components of the variation vanish.  From the $zz$- and $z\varphi_1$-components of the variation of $F_4^2$, we can see that the contributions to the boundary equations of motion dies at worst as $u^4$ as $u\to 0$.  The analysis of the two-charge solution follows similarly, and so we can proceed using \eq{Tij-one-point-function} without modification.   Allowing for $q_2\neq0$ modifies the variation of $F_{MNPQ}F^{MNPQ}$ but crucially does not introduce any leading terms in the small $u$ expansion.

Now that we have established that the variation of $F_{4}^2$ decays sufficiently fast near the AdS$_7$ boundary, we can proceed using the logic of \cite{deHaro:2000vlm} recapped above to compute $\left<T_{ij}\right>$.  To do so, we first map \eq{Two-charge-metric} to FG gauge as in \eq{Two-charge-FG-metric}, which we reproduce here for clarity
\begin{align*}
    ds^2_{\rm FG} =& \frac{L^2}{u^2}(du^2 + \hat{\alpha}_{\rm AdS}ds_{\rm AdS_5}^2+\hat{\alpha}_z dz^2)+L^2s_\theta^2\hat{\alpha}_{z\varphi_1}dzd\varphi_1 +L^2 c_\aleph^2c_\theta^2\hat{\alpha}_{z\varphi_2}dzd\varphi_2\\
    &+\frac{L^2}{4}(\hat{\alpha}_\theta d\theta^2 +s_\theta^2\hat{\alpha}_{\varphi_1}d\varphi_1^2 + c_\theta^2(\hat{\alpha}_\aleph d\aleph^2+c_\aleph^2\hat{\alpha}_{\varphi_2}d\varphi_2^2) + \hat{\alpha}_{\theta\aleph}d\theta d\aleph).
\end{align*}
The $\hat\alpha$ metric functions are given in \eq{Two-charge-FG-alphas}.  In order to put the dimensionally reduced metric in the form of \eq{KK-reduced-metric}, we then write $ds_{\rm FG}^2$ as a fluctuation around AdS$_7\times\mathds{S}^4$
\begin{align}
    ds^2 = (g^{(0)}_{\mu\nu} +h_{\mu\nu})dx^\mu dx^\nu
\end{align}
where
\begin{align}
    g^{(0)}_{\mu\nu}dx^\mu dx^\nu = \frac{L^2du^2}{u^2}+\frac{L^2}{u^2}\left(\left(1+\frac{u^2}{2} +\frac{u^4}{16}\right)ds_{\rm AdS_5}^2+\left(1-\frac{u^2}{2}+\frac{u^4}{16}\right)dz^2 \right)+\frac{L^2}{4}d\Omega_4^2.
\end{align}
Using the expressions in \eq{Two-charge-FG-alphas}, we can compute the zero modes of the fluctuations around the AdS$_7$ directions 
\begin{align}\begin{split}\label{eq:Fluctuation-7d}
        \bar{h}_7 =& -\frac{2L^2(q_1+q_2)}{15}u^4 (ds_{\rm AdS_5}^2        -5dz^2).
\end{split}
\end{align}
Similarly, the trace fluctuations on the $\mathds{S}^4$ are found to be
\begin{align}
    \varsigma = \frac{10q_2c_{2\aleph}c_{\theta}^2+5(q_2-2q_1)c_{2\aleph}+2q_1-3q_2}{8}u^4+\ldots.
\end{align}
Integrating the internal space fluctuations over the $\mathds{S}^4$ gives $\bar\varsigma=0$. The vanishing of the zero modes of the trace fluctuations means that the dimensionally reduced metric is already in FG form. The resulting stress-energy tensor one-point function is 
\begin{equation}\label{eq:unsubtractedtij}
    \langle T_{ij}\rangle\ dx^idx^j=\frac{N^3}{192\pi^3}\left[1-\frac{32}{5}(q_1+q_2)\right]\left(ds^2_{\rm AdS_5}-5dz^2\right).
\end{equation}

As it stands, \eq{unsubtractedtij} contains information associated with the holographic description of the physics of both the ambient and defect theories, and hence, the universal divergences that it contains are ambiguous.  A commonly method used to isolate quantities associated with the defect degrees of freedom is the so-called vacuum (or background) subtraction scheme.  In short, this scheme involves using a SUGRA solution that is holographically dual to the theory without a defect -- or, in some sense, containing a ``trivial'' defect.  A crucial part of this construction is that the vacuum used to regulate the theory and remove the contributions sourced by ambient degrees of freedom must have the same asymptotic behavior in the UV as the original, unregulated SUGRA solution.  For our purposes in both the immediate case of the two-charge solution and in the electrostatic solutions below, the holographic description of the large $N$ limit of ambient theory -- the $6d$ $A_{N-1}$ $\CN=(2,0)$ SCFT with no defect insertions -- is well known to be simply Ads$_7\times\mathds{S}^4$, which can be obtained from the two-charge solution by taking the $q_I\to0$ limit.  We will use this limit to compute $\langle T_{ij}^{\rm(vac)}\rangle$, which will be subtracted from \eq{unsubtractedtij} to obtain the change in the stress-energy tensor one-point function due to the presence of the defect.

Before proceeding, we note that there are limitations to using vacuum subtraction that, while not affecting any of the computations below, are nonetheless important to keep in mind.  As is alluded to above, in general it may not always be obvious how to correctly choose the background to be removed, as the notion of a ``defect'' is not always well-defined\footnote{This subtlety can be seen in familiar cases like in choosing Neumann/Robin or Dirichlet conditions for scalar fields in the presence of a boundary \cite{McAvity:1995zd} or magnetic (`t Hooft) line operators in gauge theories, see e.g. \cite{Aharony:2013hda}, as well as in more exotic contexts such as surface operators in ambient theories with unusual deformations \cite{Capuozzo:2024onf}.}. Furthermore, it is not clear that background subtraction is a justifiable scheme for computing non-universal quantities due to a lack of a systematic way to account for finite counterterms.  However, for the log-divergent quantities that we compute in this work, such subtleties can be ignored.  A more precise, alternative method for holographically computing defect quantities would involve generalizing the standard tools in holographic renormalization, but as of yet, correctly identifying a set of suitable covariant counterterms for holography with AdS submanifolds remains an open problem.

With an understanding of the utility of the background subtraction scheme in hand, we can now compute $\langle T_{ij}^{\rm(vac)}\rangle$ computed using vacuum AdS$_7\times\mathds{S}^4$, which obtain by taking $q_I\to0$ in \eq{Fluctuation-7d}.  This limit kills the fluctuations and gives the exact AdS$_7$ metric upon dimensional reduction, as expected. So, taking $q_I\to 0$ in \eq{unsubtractedtij} yields the  vacuum 1-pt function
\begin{equation}\label{eq:Vacuum-stress-tensor-one-point-function}
    \langle T_{ij}^{\rm(vac)}\rangle\ dx^idx^j=\frac{N^3}{192\pi^3}\left(ds^2_{\rm AdS_5}-5dz^2\right).
\end{equation}
Subtracting this vacuum contribution from \eq{unsubtractedtij} computes the change in the stress-energy tensor one-point function due to the introduction of the holographic dual to the field theory defect:
\begin{align}
    \Delta\left< T_{ij}\right>dx^idx^j =-\frac{N^3(q_1+q_2)}{30\pi^3}(ds_{\rm AdS_5}^2-5dz^2),
\end{align}
which recovers the results in \cite{Gutperle:2022pgw} up to subtraction of the contribution from the AdS$_7\times\mathds{S}^4$ vacuum. Using \eq{stress-tensor-one-pt-fn} we arrive at 
\begin{align}
    h_T = \frac{N^3(q_1+q_2)}{30\pi^3}
\end{align}
Thus, one of the B-type anomaly coefficients for $1/4$-BPS co-dimension 2 operators in a $6d$ $\CN=(2,0)$ $A_{N-1}$ SCFT holographically described by the two-charge solutions is found to be
\begin{align}\label{eq:Two-charge-d2}
    d_2 = -\frac{1}{6}N^3(q_1+q_2).
\end{align}

\begin{figure}[t]
    \centering
    \includegraphics[width=0.7\textwidth]{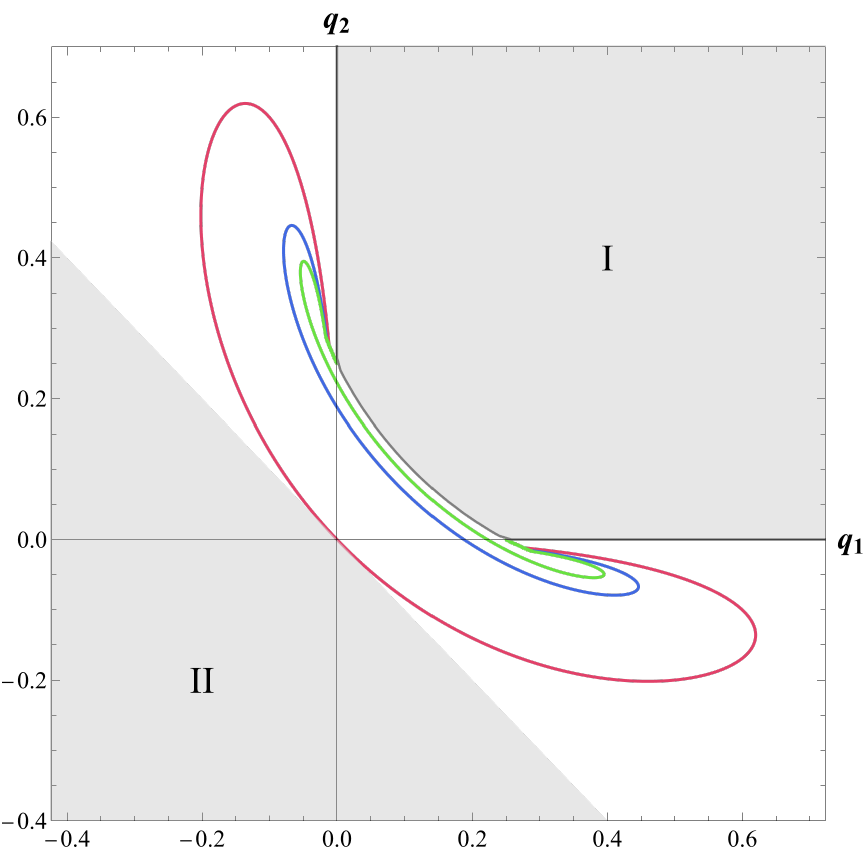}
    \caption{The solutions to the constraint in \eq{q1q2positivity} for $\hat{n}=1$ (red), $\hat{n}=2$ (blue), and $\hat{n}=3$ (green) on the $(q_1,q_2)$ plane, reproduced from \cite{Gutperle:2022pgw,Gutperle:2023yrd}. The shaded regions correspond to the two-charge configurations for which $Q(y)=0$ admits no real solutions (region I) or which violate the defect ANEC (region II).}\label{fig:Two-charge-plot}
\end{figure}

Recall that in \eq{q1q2positivity}, we found that the linear combination $q_1+q_2\geq 0$ for all $\hat{n}$. Further, we know that \eq{ANEC} implies $d_2\leq 0$, and so all of the two-charge solutions studied in \cite{Gutperle:2022pgw,Gutperle:2023yrd} are consistent with the defect ANEC.  In \fig{Two-charge-plot}, we reproduce the curves for solutions obeying \eq{q1q2positivity} as appear in \cite{Gutperle:2022pgw,Gutperle:2023yrd} together with the region excluded by consistency with defect ANEC. We see that, indeed, all of the $\hat{n}=1,\,2,\,3$ solutions lie above the line $q_1+q_2\geq0$ with only $\hat{n}=1$ saturating the bound at $q_1=q_2 =0$.

\subsection{Electrostatic solutions}\label{sec:Holographic-Tij-Electrostatic}

Prior to approaching the holographic computation of $\Delta\left<T_{ij}\right>$ for the electrostatic solutions using the methods outlined above, we again must verify that the boundary equations of motion in the dimensionally reduced geometry are unmodified by the four-form flux.  From \eq{LLM-defect-C3}, we can compute $F_4$.  For brevity, we will immediately define $r=\varrho c_\omega$ and $\eta = \varrho s_\omega$ to map \eq{LLM-defect-C3} into $(\varrho,\omega)$ coordinates on the internal space and adopt $(z,\varphi)$ using \eq{Electrostatic-z-varphi-transformation} as our angular coordinates and compute the large $\varrho$ expansion to leading order in each component
\begin{align}
 \frac{F_4}{2\k_{11}} =&\left[c_\omega^2s^3_\omega\frac{5m_3-2m_1^3}{\varrho^3} 
  \right.d\varrho\wedge dz +3c_\omega s_\omega^2m_1
     d\omega\wedge dz\\\nonumber
   &\qquad +s_\omega^3\frac{4(m_3-m_1^3)}{\varrho^3}
   d\varrho\wedge d\varphi
   +\left.c_\omega s_\omega^2\frac{6(m_1^3-m_3)}{\varrho^2}
    d\omega \wedge d\varphi \right]\wedge\ \vol{\mathds{S}^2}+\ldots
\end{align}
where we have fixed $\CC_z=-2$ following the discussion in \app{Electrostatic-FG}. 

Now, we can check the fall off of the contribution of the variation of $F_4^2$ to the equations of motion. Keeping $\CC_z = -2$ fixed and transforming into FG gauge, we find the leading behavior in the small-$u$ expansion (up to numerical factors)
\begin{align}\allowdisplaybreaks
\begin{split}
    F_{u MNP}F_u{}^{MNP}  &\sim s_{2\theta}^2 (m_1^3-m_3)^2 u^6+\ldots~,\\
    F_{z MNP}F_z{}^{MNP}&\sim  (13-5c_{2\theta})(m_1^3-m_3)^2u^8+\ldots~,\\
    F_{\varphi MNP}F_z{}^{MNP}&\sim(m_1^3-m_3)s_{\theta}^2u^4+\ldots~,\\
    F_{a MNP}F_b{}^{MNP} &\sim g_{\mathds{S}^4}+\ldots~,
\end{split}
\end{align}
where  $a,\,b$ are indices for $\mathds{S}^4$ coordinates $\{\theta, \varphi, \mathds{S}^2\}$, $g_{\mathds{S}^4}$ is the metric on the unit $\mathds{S}^4$ in $\mathds{S}^1\times \mathds{S}^2$ coordinatization.  Note that the variations in the AdS$_5$ directions vanish identically.   So, in the $u\to 0$ limit, there are no surviving contributions to the equations of motion in the dimensionally reduced geometry coming from the variation of the $F_4^2$ term. 

We can now proceed with \cite{deHaro:2000vlm}. First, we rewrite the metric in \eq{Electrostatic-FG-metric} as fluctuations around AdS$_7\times \mathds{S}^4$.  The perturbation away from AdS$_7\times \mathds{S}^4$ takes the form
\begin{align}
    \begin{split}
        h_{11} =&\ \frac{L^2}{u^2}\Big(\alpha_{\rm AdS}-1-\frac{u^2}{2}-\frac{u^4}{16}\Big)ds_{\rm AdS_5}^2 +\frac{L^2}{u^2}\Big(\alpha_{z}-1+\frac{u^2}{2}-\frac{u^4}{16}\Big)dz^2 \\
        &+\frac{L^2}{4}(\alpha_\theta -1)d\theta^2 + \frac{L^2s_\theta^2}{4}(\alpha_\varphi-1)d\varphi^2+\frac{L^2c_\theta^2}{4}(\alpha_{\mathds{S}^2}-1)d\Omega_2^2+L^2s_\theta^2\alpha_{z\varphi}dzd\varphi.
    \end{split}
\end{align}
 Fixing $\chi=-z-\varphi$ and $\beta = 2z +\varphi$, we can compute the zero modes for the AdS$_7$ part of the fluctuations, 
\begin{align}
    \bar{h}_7 = L^2\frac{m_3-m_1^3}{30m_1^3}u^4 ds_{\rm AdS_5}^2 +L^2\frac{m_1^3-m_3}{6m_1^3}u^4 dz^2 +\ldots~.
\end{align}
The trace $\mathds{S}^4$ fluctuations are found to be
\begin{equation}
    \varsigma=(1-5c_{2\theta})\frac{m_1^3-m_3}{16m_1^3}u^4-11(1-5c_{2\theta})\frac{m_1^3-m_3}{216m_1^3}u^6+\ldots~.
\end{equation}
Integrating $\varsigma$ over the $\mathds{S}^4$, we find the zero modes $\bar\varsigma =0$. The reduced geometry
\begin{align}
    g_7 =\left(1+\frac{\bar\varsigma}{5}\right)g^{(0)} + \bar{h}_7
\end{align} 
is thus already in FG form. So, the dimensionally reduced metric is
\begin{align}\label{eq:Electrostatic-7d-metric}
\begin{split}
    ds_7^2 =& \frac{L^2}{u^2}\left[du^2 +\left(1 +\frac{u^2}{2}+\frac{u^4}{16} +\frac{(m_3 -m_1^3)u^6}{30 m_1^3}\right)ds_{\rm AdS_5}^2 \right.\\
    &\hspace{0.75cm}\left.+\left(1-\frac{u^2}{2}+\frac{u^4}{16} +\frac{(m_1^3-m_3)u^6}{6m_1^3}\right)dz^2 \right],
\end{split}
\end{align}
where we have suppressed higher powers of $u$. From this expression for $ds_7^2$, we can easily read off $g_{(0)},\,g_{(2)},\,g_{(4)},$ and $g_{(6)}$.  Note if we take $n=1$ and $k_1=1$, then $m_3 = m_1^3 = N^3/8$, and so in this limit, \eq{Electrostatic-7d-metric} reduces to the exact AdS$_7$ metric, which is expected from eqs.~(\ref{eq:LLM-defect-metric}), (\ref{eq:Electrostatic-Laplace}), and (\ref{eq:n-kink}).

Proceeding with the computation in the same way as the previous subsection, we find that the holographic stress-energy tensor one-point-function takes the form
\begin{align}\label{eq:Electrostatic-stress-tensor-one-point-function}
  \langle T_{ij}\rangle\ dx^i dx^j = -\frac{N^3(3m_1^3-8m_3)}{960\pi^3m_1^3}\left(ds^2_{\rm AdS_5}-5dz^2\right).
\end{align}
Regulating this result by subtracting the AdS$_7\times\mathds{S}^4$ vacuum contribution $\langle T_{i j}^{\rm(vac)}\rangle$ in \eq{Vacuum-stress-tensor-one-point-function} produces
\begin{align}\label{eq:Electrostatic-defect-stress-tensor-one-point-function}
      \Delta \langle T_{ij}\rangle\ dx^i dx^j = -\frac{N^3(m_1^3-m_3)}{120\pi^3m_1^3}\left(ds^2_{\rm AdS_5}-5dz^2\right).
\end{align}
As a quick check, computing the trace of \eq{Electrostatic-defect-stress-tensor-one-point-function} gives $\Delta \left<T^i{}_i\right> =0$ as expected due to defect conformal symmetry. 
Comparing \eq{Electrostatic-defect-stress-tensor-one-point-function} to \eq{stress-tensor-one-pt-fn}, we find
\begin{align}\label{eq:Electrostatic-hT}
    h_T =     \frac{m_1^3-m_3}{15\pi^3}.
\end{align}
We can thus read off the defect Weyl anomaly coefficient $d_2$ from \eq{h-d2-q=2}:
\begin{subequations}
\begin{align}\label{eq:Electrostatic-d2-1}
     d_2 &=-\frac{m_1^3-m_3}{3}\\\label{eq:Electrostatic-d2-2}
     &=-\frac{1}{24}\left(N^3 - \sum_{a} \frac{N_a^3}{k_a^2}\right),
\end{align}
\end{subequations}
where in the second line we have rewritten $d_2$ in terms of the parameters $N_a$ and $k_a$ which are more suitable for comparison to field theory.  For any partition of $N=\sum_a N_a$, it is clear that $d_2\leq0$.  The upper bound $d_2=0$ is only saturated in the vacuum case $n=1,\,k_1=1$ where there is no defect. 

There is a non-trivial consistency check on the value of $d_2$ in the $n=1$ case.  As mentioned above, the $11d$ uplift of the $1/2$-BPS one-charge solutions is related to the single-kink electrostatic solutions by setting $n=1$ and $k_1 = 1/\sqrt{1-4q_1}$. Plugging these values into \eq{Electrostatic-d2-2} results in $d_2 = -N^3q_1/6$.  Checking this against the one-charge solutions found by taking $q_2\to0$ in \eq{Two-charge-d2}, we also find $d_2 = -N^3q_1/6$.  Thus, the values of $d_2$ computed in the two-charge and $n$-kink electrostatic solutions are consistent in this limit.

\section{Defect sphere EE and the defect A-type anomaly}\label{sec:Holographic-EE}

In the following subsections we will use the techniques developed in \cite{Jensen:2013lxa,Estes:2014hka} to holographically compute the defect contribution to the EE of a spherical region in the dual $6d$ $A_{N-1}$ $\CN=(2,0)$ SCFT at large $N$ for both the $1/4$-BPS two-charge and $1/2$-BPS electrostatic co-dimension $2$ defects.  Leveraging the results of the previous section and eqs.~(\ref{eq:defect-EE}) and (\ref{eq:defect-EE-anomalies}), we will be able to compute the defect A-type anomaly $a_\Sigma$. 

To facilitate the discussion below, let us briefly review some of the relevant background concepts for defect EE.  We will restrict our discussion here to the holographic duals to 6$d$ (D)SCFTs.

To start, we will need the Ryu-Takayanagi (RT) formula for holographic EE \cite{Ryu:2006bv,Ryu:2006ef,Nishioka:2009un}, which we write agnostic to the presence of a defect as
\begin{align}\label{eq:EE-formula}
    S_{\rm EE} = \frac{\CA_{\rm min}}{4G_N}.
\end{align}
The quantity $\CA_{\rm min}$ is the area of the extremal surface that minimizes the bulk area functional subject to the condition that the surface anchored at the conformal boundary of AdS$_7$ is homologous to the entangling region in the dual theory.  For our computations below, we take the entangling region in the 6$d$ SCFT at a fixed time slice to be a Euclidean 5-ball $\CB =\mathds{B}^5\hookrightarrow \mathbb{R}^5$ of radius $R$.  When we consider the theory deformed by a flat embedding of a Lorentzian defect on $\Sigma = \mathbb{R}^{1,3}$, we will take the defect to be co-original with the entangling surface such that $\pd\CB\cap \Sigma = \mathds{S}^2$ sitting along the equator of $\pd\CB$.

By including a defect in the field theory, there are subtleties that arise in directly applying \eq{defect-EE}. On the field theory side, $S_{\rm EE}$ will now have short-distance divergences near $\pd\CB$ due to highly entangled UV modes in both ambient and defect localized theories.  In the holographic description, one needs to adopt a suitable regularization scheme that isolates the defect contribution to $S_{\rm EE}$; we will use a background subtraction scheme ($S_{\rm EE}[\Sigma]-S_{\rm EE}[\emptyset]$) akin to the one used in computing the holographic stress-energy tensor one-point function.  One further complication in the holographic computation is the fact that the FG expansion is generally not globally defined, and so one must be careful to find the asymptotic form of the map to FG gauge in order to define the UV cutoff slice at fixed AdS$_7$ radius $\Lambda\gg L$.  A general formula for finding the asymptotic form of the FG transformation and cutoff slice was found in \cite{Estes:2014hka}, which we will use in the computations below.

 Since we are considering a spherical entangling region, the solution for $\CA_{\rm min}$ takes a particularly simple form; even in the presence of a defect.  It was shown in \cite{Jensen:2013lxa} that for a bulk geometry realizing the defect symmetry group $SO(2,d-\mf{d})\times SO(\mf{d})$, the relative warp factors of the AdS$_{\mf{d}+1}$ and $\mathds{S}^{\mf{d-1}}$ spaces are largely immaterial, and the logic of \cite{Casini:2011kv} can be generalized to prove \eq{defect-EE} for these backgrounds.  In the process, the authors of \cite{Jensen:2013lxa} proved that for the holographic defect spherical EE the surface $\CA_{\rm min}$ is simply a hemispherical region extending into the bulk anchored at $\CB$. For the $11d$ backgrounds corresponding to both the two-charge and electrostatic solutions that we consider, if we write the line element on the AdS$_5$ in the form
\begin{align}\label{eq:AdS5-metric-EE}
    ds_{\rm AdS_5}^2 = \frac{1}{w^2}(dw^2 - dt^2 + dr_\parallel^2 + r_\parallel^2 d\Omega_2^2)~,
\end{align}
then $\CA_{\rm min}$ is the surface $w^2 +r_\parallel^2 = R^2$.  We will exploit the simplicity of the minimal surface to great effect in the subsequent computations.

\subsection{Two-charge solutions}\label{sec:Holographic-EE-Two-charge}

To begin computing the defect spherical EE for the two-charge solutions, we need to express the area functional $\CA$ in terms of the metric functions, $\hat{f}$ in \eq{Two-charge-metric} with the AdS$_5$ factor written as in \eq{AdS5-metric-EE}. Evaluating on the extremal surface $r_\parallel^2+w^2=R^2$, we  regularize the $w$ integration by introducing a UV cutoff $\eps_w\ll 1$ and performing the integral over the angular coordinates $\phi_1,\phi_2$ and $z$ to obtain 
\begin{align}
\CA_{\rm min}[\Sigma] = 8\pi^4L^9R \int_{\eps_w}^\infty dw \frac{\sqrt{R^2-w^2}}{w^3}\CI = 4\pi^4\left(\frac{R^2}{\eps_w^2}-\log \frac{2R}{\eps_w}+\ldots\right)\CI~,
\end{align}
where we have defined the remaining integral
\begin{align}
    \CI\equiv  \int d\psi\, d\zeta \int_{y_+}^{\Lambda_y(\eps_u,\psi,\zeta)}dy\,\hat{f}_{\rm AdS}^3 f_y \sqrt{(4\hat{f}_\psi^2 \hat{f}_\zeta^2-\hat{f}_{\psi\zeta}^4)(\hat{f}_{\phi_1}^2 \hat{f}_{z\phi_2}^4 +\hat{f}_{\phi_2}^2\hat{f}_{z\phi_1}^4 - 4\hat{f}_{\phi_1}^2\hat{f}_{\phi_2}^2\hat{f}_z^2)}~.
\end{align}
Despite the initially complicated appearance of the integrand upon substituting the form of the metric functions in \eq{Two-charge-fs}, we find after a bit of algebra that the remaining integral drastically simplifies to
\begin{align}
    \CI = \frac{1}{8}\int d\psi~ d\zeta ~ c_\psi c_\zeta^2 s_\zeta\int_{y_+}^{\Lambda_y(\eps_u,\psi,\zeta)} dy~ y.
\end{align}

Using the double-cutoff prescription to compute $\CI$ as in \cite{Estes:2014hka,Gentle:2015jma}, we first map the radial coordinate $y$ to the FG coordinate $u$ leaving the remaining angular coordinates $\psi$ and $\zeta$ in their original frame. We then impose a cutoff $\eps_u\ll 1$, which induces a cutoff in large $y$, $\Lambda_y(\eps_u,\psi,\zeta)$. Recalling the asymptotic FG map in \app{Two-charge-FG} used in the previous section and recasting the FG angular coordinates $\aleph,\theta$ in terms of $\psi,\zeta$, we find that
\begin{align}\begin{split}
    \Lambda_y(\eps_u,\psi,\zeta) &= \frac{1}{\eps_u^2}+\frac{1}{2} + \frac{3-10q_1-9q_2-2q_2c_{2\psi}c_\zeta^2 +(2q_1-q_2)c_{2\zeta}}{48}\eps_u^2+\ldots~.
\end{split}
\end{align}
Evaluating the integral $\CI$ with this cutoff is straightforward, yielding
\begin{align}
    \CI = \frac{1}{24\eps_u^4} + \frac{1}{24\eps_u^2}+\frac{1}{960}(15 -16(q_1+q_2)-40y_+^2)+\ldots
\end{align}

In order to find the contributions coming from the defect, we must regulate the $\eps_u$ divergences present in $\CA_{\rm min}$.  In order to do so, we employ the same vacuum subtraction scheme as was used in computing $\Delta\left<T_{ij}\right>$ above. For the two-charge solution, the vacuum is obtained by setting $q_1=q_2=0$ and $a_1=a_2=0$, which sets $y_+^{(\text{vac})}=1$. Recomputing $\CA_{\rm min}[\emptyset]$ for the vacuum solution and subtracting it from $\CA_{\rm min}[\Sigma]$, the regulated area functional gives 
\begin{align}\label{eq:Two-charge-A-min-regulated}
    \CA_{\rm min}[\Sigma]- \CA_{\rm min}[\emptyset]=-\frac{\pi^4L^9}{30}(2q_1+2q_2+5(y_+^2- 1))\left(\frac{R^2}{\eps_w^2} -\log\frac{2R}{\eps_w}+\ldots\right)~,
\end{align}
free from $\eps_u$ divergences. 

In order to compute $a_\Sigma$ for the defect theory, we insert \eq{Two-charge-A-min-regulated} in \eq{EE-formula}.  Mapping to field theory quantities by $L^3 = 4\pi N \ell_P^3$ and $G_N = 2^4\pi^7\ell_P^9$, we can read off the coefficient of the universal part of the defect sphere EE from \eq{EE-formula}
\begin{align}
   - R\pd_R(S_{\rm EE}[\Sigma]- S_{\rm EE}[\emptyset])|_{R\to0} = -\frac{N^3}{30}(2(q_1+q_2)+5(y_+^2-1)).
\end{align}
Hence, using $d_2 = -\frac{N^3}{6}(q_1+q_2)$ derived above in \eq{defect-EE-anomalies} we find
\begin{align}\label{eq:Two-charge-a}
    a_\Sigma &=\frac{N^3}{24}(1-y_+^2).
\end{align}

One interesting consequence of this computation is that one can show that A-type anomaly of the general two-charge solution must satisfy $a_\Sigma \geq 0$. To see this more clearly, recall from \eq{q1q2positivity} that 
\begin{align}
    y_+\leq \frac{3\hat{n}+1}{4\hat{n}}\leq 1.
\end{align}
The second inequality follows from $\hat{n}\in\mathbb{N}$, and so the upper bound is saturated only for $\hat{n}=1$. Thus, for all consistent two-charge solutions, $a_\Sigma\geq0$.

\subsection{Electrostatic solutions}\label{sec:Holographic-EE-Electrostatic}

Continuing with the logic used in the previous subsection, we now turn our attention to the electrostatic solutions.  Our starting point for the computation is in transforming the metric in  \eq{LLM-defect-metric} using \eq{Electrostatic-z-varphi-transformation} and reading off the metric functions. Since only $\CC_z=-2$ gives an asymptotic form for the metric suitable for mapping into FG gauge, we fix the transformation $\chi = -z-\varphi$ and $\beta = 2z+\varphi$ and arrive at
\begin{align}\label{eq:Electrostatic-EE-transformed-metric}
   ds_{11}^2 = f_{\rm AdS}^2ds_{\rm AdS_5}^2 + f_{\mathds{S}^2}d\Omega_2^2 + f_{z}^2dz^2+f_\varphi^2d\varphi^2+f_{z\varphi}^2dzd\varphi + f_{\varrho}^2d\varrho^2+f_\omega^2d\omega^2.
\end{align}
We will also write the AdS$_5$ line element as in \eq{AdS5-metric-EE}.

 Plugging in the expression for the minimal surface, $r_\parallel^2+w^2=R^2$, into the area functional, we first integrate over the two $\mathds{S}^2$ factors as well as the angular coordinates $z\in[0,2\pi]$ and $\varphi\in[0,2\pi]$, which yields
\begin{align}
    \CA_{\rm min}[\Sigma] = 32 \pi^4 R \int  dw \frac{\sqrt{R^2-w^2}}{w^3} \CI[\Sigma]~ .
\end{align}
where
\begin{align}
    \CI[\Sigma] \equiv \int_{0}^{\pi/2} d\omega \int_{0}^{\Lambda_\varrho(\epsilon_u,\omega)}f_{\rm AdS}^3f_{\mathds{S}^2}^2f_\omega f_\varrho \sqrt{4f_z^2f_\varphi^2-f_{z\varphi}^4}~.
\end{align}
Note that we have introduced the large $\varrho$ cutoff, $\Lambda_\varrho$, that was induced by the small $u$ cutoff in FG gauge $\epsilon_u$:
\begin{align}
    \Lambda_\varrho(\epsilon_u,\omega)=&\frac{2m_1}{\epsilon_u^2}+\frac{2m_1^3s_\omega^2-(1+5c_{2\omega})m_3}{48m_1^2}\epsilon_u^2+s^2_\omega\frac{m_3-m_1^3}{36m_1^2}\epsilon_u^4
    +\ldots~.
\end{align}
Since the metric functions $f$ are independent of $w$, the $w$ integral can be performed over $[\eps_w,\infty)$, where $\e_w\ll 1$,
\begin{align}
    \CA_{\rm min}[\Sigma] = 16\pi^4\left(\frac{ R^2}{\e_w^2}-\log\frac{2R}{\e_w}+O(\e_w^0)\right)\CI[\Sigma]~.
\end{align}

Using the expressions for the metric functions in \eq{LLM-defect-metric} in terms of the potential, we find that $\CI$ can be expressed as a total derivative.  To see this more clearly, we note that in $(\varrho, \omega)$ coordinates 
\begin{align}
    \CI[\Sigma] = 64\k_{11}^{3}  \int_{0}^{\pi/2} d\omega \int_{0}^{\Lambda_\varrho(\epsilon_u,\omega)} d\varrho \, \varrho^2 c_\omega \dot{V} V^{\prime\prime}~.
\end{align} 
Switching to \((r,\eta)\) coordinates and using the Laplace equation $\ddot{V}=-r^2V^{\prime\prime}$, we arrive at
\begin{align}\label{eq:EE-I-2}
    \CI[\Sigma] = -32\k_{11}^{3}  \int_0^{\Lambda_\eta} d\eta \int_0^{\Lambda_r}dr  \, \partial_r \dot{V}^2~,
\end{align}
where we have mapped the asymptotic cutoff in $\varrho$ back to the $(r,\eta)$ frame,
\begin{align}
    \Lambda_r = \Lambda_\varrho(\eps_u,\omega)c_\omega~,\qquad \Lambda_\eta = \Lambda_\varrho(\eps_u,\omega)s_\omega~.
\end{align}

The remaining integral in $\CI$ is identical to the one found in computing the central charge for the compact electrostatic solutions in \cite{Bah:2022yjf} and again in \cite{Gutperle:2023yrd}.  For clarity, let us analyze $\CI$ in detail here. We can integrate the total derivative in \eq{EE-I-2} and find that the surviving contributions come from the boundary of the region in the $\varrho-\omega$ quarter-plane spanned by the $\eta$-axis at $\omega = \pi/2$ and the contour at fixed $\varrho=\Lambda_\varrho$ between $\omega=0$ and $\omega=\pi/2$. The integral along the $\eta$-axis can be decomposed into the regions of $\eta\in[0,\eta_n]$ and $\eta\in[\eta_n,\Lambda_{\varrho}(\eps_u,\pi/2)]$; in the latter region, the line charge density takes the form  $\lambda(\eta) = \eta + m_1$.  In all, 
\begin{align}
    \frac{\CI[\Sigma]}{32\k_{11}^3 } = \underbrace{\int_{0}^{\eta_n}d\eta \varpi(\eta)^2}_{I_1} +\underbrace{\int_{\eta_n}^{\Lambda_\varrho(\eps_u,\pi/2)}d\eta(\eta+m_1)^2}_{I_2} - \underbrace{\int_{\omega=0}^{\omega =\pi/2}\dot{V}^2\Big|_{\Lambda_r} d(\Lambda_\r(\eps_u,\omega))}_{I_3},
\end{align}
where $\dot{V}^2\Big|_{\Lambda_r}$ in $I_3$ is held at fixed $r= \Lambda_{r}$ in the integration over $\omega$.  

Let's take each of the $I$'s individually, starting with $I_2$.  Performing the integral is trivial and leads to the small $\eps_u$ expansion
\begin{align}
    I_2  =&\frac{8m_1^2}{3\eps_u^6}+\frac{4m_1^3}{\eps_u^4}+ \frac{13m_1^3+2m_3}{6\eps_u^2}+\frac{8m_3+m_1^3-18m_1^2\eta_n -18m_1 \eta_n^2 - 6\eta_n^3}{18}+\ldots~.
\end{align}

The integral $I_3$ can also be easily taken. First, we expand the integrand using the large $\varrho$ expansions of the potential in \eq{Electrostatic-asymptotic-V}. Then after computing $d\Lambda_r(\eps_u,\omega)$, we expand in small $\eps_u$ and integrate term-by-term in $\omega\in[0,\pi/2]$, which gives
\begin{align}
    I_3 =\frac{8m_1^3}{3\eps_u^6}+\frac{8m_1^3}{3\eps_u^4}+\frac{5m_1^3+2m_3}{6\eps_u^2}+\frac{m_1^3+14m_3}{45}+\ldots~.
\end{align}
Combining $I_2$ and $I_3$, we see
\begin{align}
I_2-I_3  = \frac{4m_1^3}{3\eps_u^4}+\frac{4m_1^3}{3\eps_u^2}+ \frac{4m_3 +m_1^3-10\eta_n(\eta_n^2+3\eta_n m_1 + 3m_1^2)}{30}+\ldots~. 
\end{align}

Lastly, we need to take care of the integral $I_1$.  To do so, we break up the the integral over $\eta\in[0,\eta_n]$ into a sum over the intervals $[\eta_a,\eta_{a+1}]$ for $a=0,\ldots,n-1$ with $\eta_0=0$. Then, using $\varpi_a = p_{a+1}\eta +\delta_{a+1}$ over each interval we find
\begin{align}
    I_1  =\frac{1}{3}\sum_{a=0}^{n-1}\left(p_{a+1}^2(\eta_{a+1}^3-\eta_a^3) +3\delta_{a+1}p_{a+1}(\eta_{a+1}^2-\eta_{a}^2)+3 \delta_{a+1}^2(\eta_{a+1}-\eta_a)\right).
\end{align}
Combining everything we get
\begin{align}\begin{split}\label{eq:Electrostatic-CI-1}
    \frac{\CI[\Sigma]}{32\k_{11}^3} =&\ \frac{4m_1^3}{3\eps_u^4}+\frac{4m_1^3}{3\eps_u^2}+\frac{4m_3+m_1^3}{30}+ \frac{1}{3}\sum_{a=0}^n(p_{a+1}^2\eta_{a+1}^3-\eta_a^3)  \\
    &+\sum_{a=0}^n\delta_{a+1}p_{a+1}(\eta_{a+1}^2-\eta_a^2) +\sum_{a=0}^n\delta_{a+1}^2(\eta_{a+1}-\eta_a),
    \end{split}
\end{align}
where we slightly abuse the notation by setting $\eta_{n+1}=0$ in this sum to make the expressions a bit more compact.

The $\eps_u$ divergences in $\CI[\Sigma]$ need to be regulated.  We again adopt the background subtraction scheme as before, where the background vacuum AdS$_7\times\mathds{S}^4$ solution is obtained by taking $n=1$ and $k_1=1$. Taking this limit in \eq{Electrostatic-CI-1} yields
\begin{align}
    \frac{\CI[\emptyset]}{32\k_{11}^3} =\frac{4m_1^3}{3\eps_u^4}+\frac{4m_1^3}{3\eps_u^2}-\frac{5m_1^3}{6}+\ldots~.
\end{align}
We then arrive at the expression for the regulated $\CI$: 
\begin{align}\label{eq:CI-regulated}
\begin{split}
    \frac{\CI[\Sigma] - \CI[\emptyset]}{32\k_{11}^3} = &\frac{2m_3+13m_1^3}{15}+ \frac{1}{3}\sum_{a=0}^np_{a+1}^2(\eta_{a+1}^3-\eta_a^3) +\sum_{a=0}^n\delta_{a+1}p_{a+1}(\eta_{a+1}^2-\eta_a^2) \\
    &+\sum_{a=0}^n\delta_{a+1}^2(\eta_{a+1}-\eta_a)),
    \end{split}
\end{align}
which recovers the result of the integral for the non-compact electrostatic solutions in \cite{Gutperle:2023yrd}. Thus, the regulated minimal area is given by
\begin{align}\label{eq:Electrostatic-regulated-A}
\CA_{\rm min}[\Sigma]-\CA_{\rm min}[\emptyset] = 2^9\pi^4\k_{11}^3\left(\frac{R^2}{\eps_w^2}-\log\frac{2R}{\eps_w}+O(1)\right)(\CI[\Sigma]-\CI[\emptyset]).
\end{align}

Proceeding with the computation of $a_\Sigma$, we feed \eq{Electrostatic-regulated-A} in \eq{EE-formula} to get $S_{\rm EE}$. Computing the log derivative with respect to $R$ of the regularized minimal area functional at $R=0$ gives the universal part of defect entanglement entropy 
\begin{align}\label{eq:Electrostatic-SEE}
\begin{split}
    R\pd_R(S_{\rm EE}[\Sigma]-S_{\rm EE}[\emptyset]) =& -(\CI[\Sigma]-\CI[\emptyset]),
    \end{split}
\end{align}
where we mapped to the field theory variables using $G_N^{(11)} =2^{13} \pi^4\k_{11}^3$ and $\k_{11}= L^3/8N$.  Using \eq{defect-EE-anomalies} we can read off the A-type anomaly coefficient using $d_2 =-\frac{1}{3}(m_1^3-m_3)$
\begin{align}\begin{split}\label{eq:Electrostatic-a-3}
 a_\Sigma = \frac{\left(\sum_{a=1}^nk_a\eta_a\right)^3}{4}+ \frac{1}{12}\sum_{a=0}^n(p_{a+1}^2(\eta_{a+1}^3-\eta_a^3) +3\delta_{a+1}p_{a+1}(\eta_{a+1}^2-\eta_a^2) +3\delta_{a+1}^2(\eta_{a+1}-\eta_a)).
\end{split}\end{align}
 Recall that the $\eta_a$ are ordered by $0=\eta_0<\eta_1<\ldots<\eta_n$, and so $(\eta_{a+1}^j - \eta_a^j)>0$ for any $j\in \mathbb{N}$ and for all $a$.  Further, the orbifold parameters are non-negative $k_a\in\mathbb{N}$, and so by definition are the $p_{a}$, and in addition $2\delta_{a}\in\mathbb{N}$.  Hence, we see that $a_\Sigma\geq 0$. Note that the inequality is saturated at $n=k_1=1$ i.e. $a_\Sigma = 0$, which is expected since this line charge density configuration corresponds to having no defect.

For completeness, we can rewrite $a_\Sigma$ in terms of the ranks, $N_a$, of the factors in the Levi subalgebra $\mf{l}\subset A_{N-1}$ and their associated monopole charges, $k_a$,
\begin{align}
  a_\Sigma =  \frac{N^3}{32} -\frac{1}{96}\sum_{a=1}^n\left(\frac{1+2k_a}{k_a^2}N_a^3 +\sum_{b=a+1}^nN_ak_b\left(\frac{N_a^2}{k_a^2}+3\frac{N_b^2}{k_b^2}\right)\right).
   \end{align}
While the definite sign of $a_\Sigma$ is a bit less clear in terms of the gauge algebra data, it is nonetheless non-negative following from \eq{Electrostatic-a-3}.

   As we mentioned toward the end of \sn{Holographic-Tij-Electrostatic}, there is a non-trivial consistency check of our results in \eq{Electrostatic-a-3} from the comparison to the one-charge ($q_2\to0$) solutions.  Setting $n\to1$ and $k_1\to1/\sqrt{1-4q_1}$ in \eq{Electrostatic-a-3} results in
   \begin{align}\label{eq:Electrostatic-a-q1}
       a_{\Sigma}\big|_{n=1}= \frac{N^3}{48}\left(1+2q_1-\sqrt{1-4q_1}\right).
   \end{align}
    Looking back to the computation of a$_\Sigma$ for the two-charge solutions, we need the largest root of $Q(y)$ with $q_2\to0$, which is simply $y_+(q_1) = \frac{1}{2}(1 +\sqrt{1-4q_1})$.  Plugging $y_+(q_1)$ into \eq{Two-charge-a} exactly matches \eq{Electrostatic-a-q1}.

We now compare $a_\Sigma$ to the computations of the `defect central charge' for these solutions. The `defect central charge' was computed in \cite{Gutperle:2023yrd} using the standard formula for the central charge $c_{\rm 4d}$ of \textit{standalone} $4d$ $\CN=2$ SCFTs at large $N$ holographically dual to AdS$_5$ solutions in M-theory \cite{Gauntlett:2006ai}
\begin{align}\label{eq:Gauntlett-central-charge}
    c_{\rm 4d} = \frac{2^5\pi^3\k_{11}^3}{(2\pi\ell_P)^9}\int_{\CM_6} \left(\frac{\dot{V}\sigma}{2V^{\prime\prime}}\right)^{\frac{3}{2}},
\end{align} 
which applies to $11d$ metrics of the form
\begin{align}
    ds_{11}^2 = \left(\frac{\k_{11}^2\dot{V}\sigma}{2V^{\prime\prime}}\right)^{\frac{1}{3}}(ds_{\rm AdS_5}^2+ds_{\CM_6}^2).
\end{align}
 This formula had been used to find the holographic central charge dual to electrostatic solutions with compact internal space engineering irregular punctures \cite{Couzens:2022yjl, Bah:2022yjf}.   Despite the integrals in eqs.~(\ref{eq:CI-regulated}) and (\ref{eq:Gauntlett-central-charge}) having the same form, the crucial difference is in the interpretation of the result: the relative difference between $a_\Sigma$ and $c_{\rm 4d}$ is a factor of $-2d_2/5$. 
 
 Lastly, while monotonicity of the universal part of the defect sphere EE has yet to be tested for $4d$ DCFTs, in the case of a co-dimension 4 Wilson surface in $6d$ SCFTs the universal defect contribution to the sphere EE does not behave monotonically under defect RG flows (see e.g. \cite{Rodgers:2018mvq}). Due to the relative sign in $\Delta S_{\rm EE}$ and the fact that only $a_\Sigma$ is known to obey a weak defect $a$-theorem\footnote{The recent entropic proof in \cite{Casini:2023kyj} of the irreversibility of defect RG flows in addition to the dilaton effective action methods (\'a la \cite{Komargodski:2011vj}) in \cite{Wang:2021mdq} have firmly established the existence of at least a weak defect $a$-theorem.}, it is expected that \eq{Electrostatic-SEE} is not a monotone along defect RG flows.

\section{On-shell action}\label{sec:On-shell-action}

Given a solution to the SUGRA equations of motion, one of the most basic quantities that one can compute is the on-shell action.  Holographically, the on-shell action is mapped to the free energy of the theory, and so with an even dimensional spherical boundary, it has universal divergences related to anomalies.

In this section, we evaluate the on-shell action for the $11d$ uplift of the two-charge solutions. This will facilitate a comparison to the same quantity computed in the realization as a domain wall in $7d$ $\CN=4$ gauged SUGRA, which was previously found in \cite{Gutperle:2022pgw}. Crucially, by employing the equations of motion, the $11d$ action can be recast as a boundary integral. Then by fixing a gauge where $C_6$ has vanishing components along cycles, e.g. the ${dz \wedge \Upsilon_{\rm AdS_5}}$ cycle, that vanish at $y=y_+$, the regulated on-shell action computed in the background subtraction scheme is then symmetric in the $q_I$, as expected from the uplift of the $7d$ theory. Interestingly, the $11d$ result does not match the on-shell action computed in \cite{Gutperle:2022pgw} for the $7d$ domain wall. We argue the discrepancy is related to the change of variables defined in \eq{Two-charge-new-phi}.  On the $11d$ side, the change of variables is induced by the transformation to FG gauge.  In the $7d$ theory, this corresponds to making a large gauge transformation, which can be seen in the uplift given in \eq{Two-charge-uplift}.  Interestingly, the choice of FG gauge seems to correspond to a singular gauge in the $7d$ theory.  We note that the change of coordinates mixes spatial rotations around the defect with the R-symmetry, which is specific to co-dimension two defects.

Moving on, the starting point for computing the on-shell action for the two-charge  solutions in \eq{Two-charge-uplift} is the bosonic part of the $11d$ SUGRA action
\begin{align}\label{eq:SUGRA-action-11d}
    S=\frac{1}{16\pi G_{N}^{(11)}}\int_{\mathcal{M}}d^{11}x\ \sqrt{-g_{11}}\left(\CR-\frac{1}{48}F_{MNPQ}F^{MNPQ}\right)+\frac{1}{8\pi G_{N}^{(11)}}\int_{\partial\mathcal{M}}K \Upsilon_{\pd\CM}+S_{\text{CS}},
\end{align}
where $\Upsilon_{\pd\CM}$ is the natural volume form associated to the metric induced on the boundary $\partial\mathcal{M}$, while $K$ is the trace of the boundary extrinsic curvature $K_{MN}=-\frac{1}{2}(\nabla_M \n_N+\nabla_N \n_M)$ with $\n_M$ denoting the components of the outward-pointing normal vector to $\partial\mathcal{M}$ and where capital Latin indices $M,\,N\in\{0,\ldots, 10\}$. Using the equations of motion for the $11d$ metric we can write the bulk term as
\begin{align}
    \sqrt{-g_{11}}\left(\CR-\frac{1}{48}F_{MNPQ}F^{MNPQ}\right) d^{11}x= -\frac{1}{3}F_4\wedge\star F_4.
\end{align}
Note that for this particular solution, the four-form flux obeys the equation
\begin{equation}
    d\star F_4=0,
\end{equation}
and consequently the Chern-Simons term $S_{\text{CS}}$ vanishes.  As a further consequence of the equations of motion for the four-form flux, we can freely exchange $\star F_4$ for $dC_{6}$, which due to $C_6$ being better behaved will make the following computation a bit easier. Using this fact and the bulk equations of motion, the bulk integrand can be expressed as a total derivative. Thus, the on-shell action can be written as a boundary integral
\begin{align}\label{eq:On-shell-action-0}
    S_{\text{OS}}    &=\frac{1}{16\pi G_{N}^{(11)}}\int_{\partial\mathcal{M}}\left(2K \Upsilon_{\pd\CM}-\frac{1}{3}F_4\wedge C_6\right) =:S_{\rm{OS},\rm{GHY}}+S_{\rm{OS},\rm{bulk}}.
\end{align}
The particular solutions we are interested in are asymptotically locally AdS$_7\times \mathds{S}^4$. So, in order to regularize the boundary integral, we first map the metric into FG form as in \eq{Two-charge-FG-metric} using the explicit asymptotic coordinate transformation derived in \eq{Two-charge-FG-transformation}. That is, we will define a regulating hypersurface at $u=\epsilon_u$ that will become $\partial\mathcal{M}$ as we take $\epsilon_u\to 0$. Note that due to the presence of an AdS$_5$ factor, an additional regularization procedure will have to be applied, which we will address later.

Before beginning the computation in earnest, we will need the asymptotic $u\ll 1$ expansions of $F_4$ and $C_6$.  First, we compute $C_6$ from \eq{Two-charge-potential}, which yields 
\begin{align}
    C_6&=L^6\Biggl\{\frac{1}{2} q_2 c_{\zeta }^2 c_{\psi }^2d\phi_2+\frac{1}{2} q_1 s_{\zeta }^2d\phi_1+\Biggl[y(y^2+q_2)-\frac{c_\zeta^2}{2y}\left(q_2 c_{2 \psi } \left(y \left(y-a_2-1\right)+q_1\right)\right.\nonumber\\&\qquad\quad\left.+2 q_1 y \left(a_1-y+1\right)+q_2 y \left(y-a_2-1\right)-q_2 q_1\right)\Biggr]dz\Biggr\}\wedge\Upsilon_{\rm{AdS}_5}.
\end{align}
We can then use the residual gauge freedom to shift $C_6\mapsto C_6+d\Lambda_5=:\tilde{C}_6$ such that $\tilde{C}_6$ has no surviving $dz\wedge \Upsilon_{\rm AdS_5}$ components at $y=y_+$.  At $y=y_+$, we can use the values for $a_I$ determined from $A_I(y_+)=0$ to show 
\begin{align}
    C_6 (y_+) &=L^6\left\{\frac{1}{2} q_2 c_{\zeta }^2 c_{\psi }^2d\phi_2+\frac{1}{2} q_1 s_{\zeta }^2d\phi_1+y_+H_2(y_+)dz\right\}\wedge\Upsilon_{\rm{AdS}_5},
\end{align}
where the terms in $\Upsilon_{\text{AdS}_5}\wedge dz$ depending on the angular coordinates vanish due to a common factor of $Q(y_+)$ appearing in their coefficients.  By demanding that the $\Upsilon_{\text{AdS}_5}\wedge dz$ part of $C_6$ vanishes at $y=y_+$, we find the appropriate gauge transformation to be
\begin{align}
    \Lambda_5=&-z L^6 y_+ H_2(y_+)\Upsilon_{\rm{AdS}_5}.
\end{align}
To implement the FG cutoff, we change coordinates to $\phi_I\rightarrow\varphi_I$ as defined in \eq{Two-charge-new-phi}.  This modifies the vanishing condition for $C_6$ at $y=y_+$, resulting in a shifted gauge transformation with the shift given by
\begin{align}
 \Lambda_5 \rightarrow \Lambda_5 + z L^6 c_{\zeta }^2 c_{\psi }^2 a_2 q_2 + z L^6 s_{\zeta }^2 a_1 q_1.
\end{align}
Using this gauge transformation we find the asymptotic expansion of $\tilde{C}_6$ to be
\begin{align}\label{eq:Two-charge-C6-FG}
    \tilde{C}_6 &= L^6\left[\left(\frac{1}{u^6}+\frac{3}{2u^4}-\frac{1}{16u^2}(2q_1-3(5+q_2)+10q_2c_{2\aleph}c_{\theta}^2+5(q_2-2q_1)c_{2\theta})\right)dz
    \right]\wedge\Upsilon_{\rm{AdS}_5}+\ldots~.
\end{align}
It is useful to note that the $\Upsilon_{\text{AdS}_5}\wedge dz$ component of $\tilde{C}_6$ does not change under the coordinate transformation $\phi_I\rightarrow\varphi_I$, once the shift in the gauge transformation is taken into account.

Next, we need to find $F_4$, which we can easily compute from \eq{Two-charge-potential}. We then map into FG coordinates, fix $\hat{g} = 2$, and expand in small $u$.  Keeping the most relevant singular terms, we find
    \begin{align}
        \label{eq:Two-charge-F4}\nonumber
        F_4&=\frac{L^3}{8}\Biggl\{\left[3c_{\theta }^2 s_{\theta }d\varphi_1\wedge d\theta+\frac{c_{\theta }^3}{2} \left(5 s_{\theta }^2 \left(2q_1-q_2 c_{2 \aleph }-q_2\right)du\wedge d\varphi_1+16 q_1 du\wedge dz\right)u^3\right]\wedge\Upsilon_{\mathds{S}^2}\\
        &\qquad\qquad+\frac{s_{\theta }|c_\aleph|}{2} du\wedge d\varphi_1\wedge \left(5 q_2c_{\theta }^2 s_{2\aleph }d\theta\wedge d\varphi_2 +8 q_2 dz\wedge\left(\frac{2s_\aleph}{c_\aleph} d\theta-s_{2 \theta } d\aleph \right)\right)u^3\Biggr\}+\ldots~.
    \end{align}
Now that we have the asymptotics of the metric, $\tilde{C}_6$, and $F_4$, we are in position to compute the on-shell action for the two-charge solutions. To begin, we first examine the Gibbons-Hawking-York (GHY) term. We note that after mapping to FG coordinates as in \eq{Two-charge-FG-metric}, the volume form on the regulating cutoff slice at $u=\eps_u$  can be easily seen to have small $\eps_u$ expansion 
\begin{align}
    \hspace{-0.85cm}\Upsilon_{\pd\CM}=\frac{L^{10}}{16}\left(\frac{1}{\epsilon_u^6}+\frac{1}{\epsilon_u^4}+\frac{5}{16\epsilon_u^2}+\frac{5\left(5 c_{2\theta} (q_2-2 q_1)+2 q_1+q_2(10  c_\theta^2 c_{2\aleph}-3)\right)}{432} \right)\Upsilon_{\text{AdS}_5}\wedge dz\wedge \Upsilon_{\mathds{S}^4}+\ldots~.
\end{align}
where we denote $\Upsilon_{\mathds{S}^4} :=  \lvert c_{\aleph}\rvert c^2_\theta s_\theta d\phi_1\wedge d\phi_2\wedge d\theta\wedge d\aleph$.
A quick calculation also shows the trace of the extrinsic curvature on the cutoff slice to be given by
\begin{align}
    K=-\frac{6}{L}+\frac{2\eps_u^2}{L}-\frac{3\eps_u^4}{4L}+\frac{\left(25 c_{2\theta}(q_2-2 q_1)+10 q_1+50 q_2 c_\theta^2c_{2\aleph}-15 q_2+9\right)\eps_u^6 }{72 L}+\ldots,
\end{align}
where we have dropped terms at $O(\eps_u^8)$ that depend on the charges but do not contribute to the final result as $\eps_u\to0$.  Thus, we find 
\begin{align}\label{eq:Two-charge-on-shell-action-GHY}
    S_{\text{OS},\text{GHY}}=-\vol{\text{AdS}_5}\frac{\pi^2 L^9}{8G_N^{(11)}}\left(\frac{2}{\epsilon_u^6}+\frac{4}{3\epsilon_u^4}+\frac{5}{24\epsilon_u^2}\right)+\ldots~.
\end{align}
Note that despite $K$ and $\Upsilon_{\pd\CM}$ containing non-trivial dependence on the charges, the end result in \eq{Two-charge-on-shell-action-GHY} is independent of the charges to $O(\eps_u^{0})$, and the $\eps_u^0$ part of the GHY term explicitly vanishes. Moving on to find $S_{\rm OS, bulk}$, using eqs.~(\ref{eq:Two-charge-C6-FG}) and (\ref{eq:Two-charge-F4}), pulling back on to the $u=\eps_u$ hypersurface, and combining with \eq{Two-charge-on-shell-action-GHY}, we arrive at 
 \begin{align}\label{eq:Two-charge-OS-bulk}
   S_{\rm OS,bulk}=&-\vol{\text{AdS}_5}\frac{\pi^2 L^9}{16 G_N^{(11)}}\left(\frac{2}{3 \epsilon_u^6} + \frac{1}{\epsilon_u^4} + \frac{5}{8 \epsilon_u^2}- \frac{
    2 q_1 (q_2 + y_+ (2 + 3 y_+))}{15 y_+}\right.\\\nonumber
    & \left.\hspace{4cm}-\frac{ 32 q_2 + 48 q_2 y_+ + 80 y_+^3-25 }{120}\right)+\ldots~.
\end{align}

Thus, combining eqs.~(\ref{eq:Two-charge-on-shell-action-GHY}) and (\ref{eq:Two-charge-OS-bulk}) and subtracting of the on-shell action for the AdS$_7\times \mathds{S}^4$ vacuum in \eq{Vacuum-on-shell-action}, which is recovered by setting $q_I = a_I = 0$ and $y_+ = 1$, the full regulated on-shell action is 
\begin{align}
    S_{\rm OS}-S_{\rm OS}^{\rm (vac)} &=\frac{\vol{\text{AdS}_5} \pi^2 L^9}{120 y_+ G^{(11)}_N}
    \left(q_1q_2 + (q_1+q_2)y_+(2+3y_+)+5y_+(y_+^3 -1)\right)\nonumber\\
    &= - \frac{\vol{\text{AdS}_5} \pi^2 L^9}{24 G^{(11)}_N} \left(1 - y_+^2 - \frac{1}{5} \left( \frac{2 q_1^2}{q_1 + y_+^2} + \frac{2 q_2^2}{q_2 + y_+^2} \right) \right)\nonumber\\
    &= - \frac{\vol{\text{AdS}_5} \pi^2 L^9}{24 G^{(11)}_N} \left(1 - y_+^2 - \frac{1}{5} ( 2 a_1 q_1 + 2 a_2 q_2 )\right).
\end{align}
 Note that choosing a different $\Lambda_5$ while maintaining regularity at $y=y_+$ does not change the final result.
Further, using the form of the regulated AdS$_5$ volume in \app{AdS5-volume}, the log divergent part of the on-shell action for the two-charge solutions is given by
\begin{align}\label{eq:Two-charge-on-shell-action}\nonumber
    S_{\rm OS}^{\rm(ren)}\big|_{\log} &=-\frac{N^3}{1920 y_+}
     \left(q_1q_2 + (q_1+q_2)y_+(2+3y_+)+5y_+(y_+^3 -1)\right)\\
     &=\frac{N^3(4q_1q_2 -2(q_1+q_2)y_+(1-y_+)+5y_+(1-y_+^2))}{1920y_+}~.
\end{align}

Before comparing to the $7d$ result, we first consider what would happen if we were to perform the computation without making the coordinate transformation in \eq{Two-charge-new-phi}.  In particular, we consider
\begin{align}\label{eq:Two-charge-new-phi-alt}
    \varphi_I = \tilde \varphi_I + 2 a_I n_I z.
\end{align}
The bulk integrand $F_4 \wedge C_6$ picks up new cross terms which are absent when $n_1 = n_2 = 0$.  The remaining terms are unmodified due to the regularity condition imposed on $C_6$ at $y = y_+$. The new terms are
\begin{align}
    \frac{3L^9}{8}\left(n_2 q_2 a_2 c_\zeta^2c_\psi^2 +n_1a_1q_1 s_\zeta^2 \right)\Upsilon_{\mathds{S}^4}\wedge \Upsilon_{\rm{AdS}_5}\wedge dz+\ldots\subset F_4 \wedge C_6~.
\end{align}
Integrating these along the boundary changes the on-shell action as follows,
\begin{align}\label{eq:11d-7d-OS-comparison}
S_{\rm OS} \longmapsto S_{\rm OS} - \frac{\vol{\text{AdS}_5} \pi^2 L^9}{24 G^{(11)}_N} \frac{1}{5} (2 n_1 a_1 q_1 + 2 n_2 a_2 q_2 )~.
\end{align}
In particular, choosing $n_1 = n_2 = 1$ reproduces the $7d$ result given in \cite{Gutperle:2022pgw}\footnote{\label{foot:discrepancy}To be clear, there is a discrepancy in the normalization between the expression in \eq{11d-7d-OS-comparison} and that found in eq.~(4.7) of \cite{Gutperle:2022pgw}.  Our expression normalizes the sign convention in the computation of the on-shell action between uplifted $11d$ solution and the $7d$ domain-wall description and fixes missing factors of $2\pi$, coming from the integral over $z$, and $L^5$.}  and corresponds to using the original coordinates $\phi_I$.  Examining the uplift defined in \eq{Two-charge-uplift}, we can see that the coordinate transformations given by \eq{Two-charge-new-phi-alt} and \eq{Two-charge-new-phi} correspond to large gauge transformations in the $7d$ description.  Thus from the $7d$ point of view, the choice of FG gauge, with $n_1 = n_2 = 0$ corresponds to a singular gauge choice.  Uplifts of lower dimensional solutions to higher dimensional ones are not necessarily unique, and it is possible that there are other $11d$ geometries which correspond to other gauge choices, such as one with $n_1 = n_2 = 1$.

\section{Discussion}\label{sec:Discussion}

In this work, we have analyzed solutions in $11d$ SUGRA that holographically describe $1/4-$ and $1/2-$BPS co-dimension $2$ defects in the $6d$ $A_{N-1}$ $\CN=(2,0)$ SCFT at large $N$. 

Our holographic computations of the defect contribution to the one-point function of the stress energy tensor have revealed simple expressions for the defect Weyl anomaly coefficient $d_2$ in \sn{Holographic-Tij}. For the $1/4$-BPS two-charge solutions specified by charges $q_1,\,q_2$, we have found that $d_2\propto N^3(q_1+q_2)$. For the $1/2$-BPS electrostatic solutions determined by a potential solving a Laplace-type equation with moments $m_j$, $d_2\propto (m_1^3-m_3)\propto N^3 - \sum_{a}N_a^3$ where $N = \sum_{a}N_a$.   Using the $4d$ form of the defect ANEC, which states $d_2\leq 0$, we have demonstrated that all of the allowed two-charge solutions found in \cite{Gutperle:2022pgw} and the electrostatic solutions in \cite{Gutperle:2023yrd} obey the bound and are thus consistent with this known defect energy condition \cite{Casini:2023kyj}.  We were also able to compare against a similar computation for the two-charge solutions done in $7d$ $\CN=4$ gauged SUGRA, and found an agreement with $\left<T_{ij}\right>$ in \cite{Gutperle:2022pgw}.

In \sn{Holographic-EE}, we used the tools developed in \cite{Jensen:2013lxa,Estes:2014hka} to holographically compute the contribution of flat, co-dimension 2 defects to the EE of a spherical region in the dual field theory.  By isolating the universal, log-divergent part of the defect sphere EE, we were able to find closed form expressions for the A-type anomaly $a_\Sigma$ for both defect systems considered.  Since we know that the universal part of the defect sphere EE, is a linear combination of $a_\Sigma$ and $d_2$ as in \eq{defect-EE}, by combining $\Delta\left<T_{ij}\right>$ and $\Delta S_{\rm EE}$, we have a direct computation of $a_\Sigma$: for the two-charge solutions we found $a_\Sigma\propto N^3 (1-y_+^2)$ where $y_+$ is the largest root of the quartic polynomial in \eq{Two-charge-Q}, while $a_\Sigma$ for the electrostatic solutions in \eq{Electrostatic-a-3} is a complicated function of the data of the line charge distribution that specifies the solution.  For the electrostatic solutions, we have shown that the computation of the holographic `central charge' in \cite{Gutperle:2023yrd} is proportional to the universal part of the defect sphere EE.  Further, we were able to show that the complicated sum over line charge density data that appears in $a_\Sigma$ is the same sum that determines the large $N$ `central charge' $c_{\rm 4d}(=a_{\rm 4d})$ for the \textit{compact} electrostatic solutions describing $4d$ $\CN=2$ SCFTs; the important difference is that the defect $a_\Sigma$ has an additional contribution of $N^3/32$. In both classes of defects, we have also shown that $a_\Sigma\geq0$, where the inequality is only saturated for a trivial defect.

Curiously, in \sn{On-shell-action}, we showed that the holographically renormalized on-shell action for the $11d$ uplift of the two-charge solutions using the full form of the radial cutoff in FG gauge and found that the log divergent part of the action cannot be written in terms of either $a_\Sigma$, as was expected from the same computation done in $7d$ gauged SUGRA description of the two-charge defects \cite{Gutperle:2022pgw}. We ultimately identified the source of this discrepancy in the parametrization of angular variables, $\phi_I$ versus $\varphi_I$, the bulk integrand. The map to FG gauge results in a redefinition of $\phi_I\to\varphi_I$ which mixes the $U(1)$ normal bundle rotations around the defect and the $U(1)$ R-symmetry. While the uplifted $11d$ SUGRA description is perfectly regular after the map to $\varphi_I$, the $7d$ gauged SUGRA picture sees this redefinition as a large gauge transformation resulting in a singular gauge. At the level of the on-shell action in the original $\phi_I$ coordinates, \eq{Two-charge-on-shell-action} picks up extra terms which, accounting for the normalization discussed in footnote \ref{foot:discrepancy}, recovers the $7d$ results in \cite{Gutperle:2022pgw}.

 With the holographic predictions for $a_\Sigma$ and $d_2$ in hand, let us compare to results in the field theory at large $N$. We will focus entirely on the $1/2$-BPS electrostatic solutions in the following comparisons.

\subsubsection*{Defect supersymmetric Casimir energy}

In ordinary $4d$ SCFTs with R-symmetry placed on $\mathds{S}^1_\beta\times\mathds{S}^3$, the supersymmetric localized partition function can be decomposed as a product of an exponential prefactor multiplying the superconformal index
\begin{align}
\CZ_{\mathds{S}^1_\beta\times\mathds{S}^3} = e^{-\beta E_C}\CI.
\end{align}
The supersymmetric Casimir energy (SCE), $E_C$, can be expressed in terms of the conformal anomalies $a$ and $c$ \cite{Kim:2012ava,Assel:2014paa} of the theory, the equivariant integral of the anomaly polynomial \cite{Bobev:2015kza}, or `t Hooft anomalies \cite{Closset:2019ucb}. Given the results in \cite{Bullimore:2014upa} for the localized partition functions a $1/2$-BPS co-dimension 2 defect in a $6d$ $\CN=(2,0)$  $A_{N-1}$ SCFT labelled by $\vartheta$ wrapping $\Sigma=\mathds{S}^1_\beta\times\mathds{S}^3\subset \mathds{S}^1_\beta\times\mathds{S}^5$, it was conjectured in \cite{Chalabi:2020iie} that the change in the exponential prefactor due to the introduction of the defect was in fact the defect SCE and could be related to defect conformal anomalies\footnote{Evidence for a version of this conjecture for $\mf{d}=2$ defects gathered from studying various examples appeared to support the claim, and in \cite{Meneghelli:2022gps}, a relation between the SCE and $h_T$ was established using the chiral algebra description of the defect insertion, which gives a much stronger argument for $E_C$ being controlled by $d_2$. We thank Maxime Tr\'epanier for pointing out the chiral algebra proof  in \cite{Meneghelli:2022gps} to us.}. Now that we have holographic predictions for two defect anomalies, we can look for a superficial match to this field theory quantity.

 As a very brief overview, we start the comparison by putting the ambient theory on the squashed $\mathds{S}^1_\beta \times \mathds{S}^5_{\mf{b}}$ and reducing along the $\mathds{S}^1$ factor. The localized partition function of the $6d$ $\CN=(2,0)$  $A_{N-1}$ SCFT in the unrefined limit becomes the partition function of $5d$ $\CN=2$ $U(N)$ super-Yang-Mills theory on $\mathds{S}^5_{\mf{b}}$, which determines the ambient SCE 
 \begin{align}
     E_C[\emptyset] \equiv \frac{\mf{c}}{24}~,\qquad \text{where} \qquad\mf{c} = N(N^2-1)(\mf{b}+\mf{b}^{-1})^2+N-1.
 \end{align}
 The quantity $\mf{c}$ in this picture is the central charge of the $2d$ $W_N$-algebra on the plane orthogonal to the directions that defect will eventually wrap \cite{Beem:2014kka,Bullimore:2014upa}. The introduction of a co-dimension 2 defect breaks the gauge algebra to the Levi subalgebra $\mf{l} = \mf{s}\left[\bigoplus_{a=1}^n\mf{u}(N_a)\right]$.   The most general $1/2$-BPS defect configuration allows for monodromy parameters $\vec{\mf{w}} = (\mf{w}_1,\ldots,\mf{w}_n)$ for the Levi factors.  The change in the SCE due to introducing the defect along $\Sigma$ labelled by $\vartheta:\mf{sl}(2)\to\mf{g}$ with monodromy parameters $\vec{\mf{w}}$ was found to be given by \cite{Bullimore:2014upa,Chalabi:2020iie}
 \begin{align}\label{eq:Defect-SCE}
      E_C[\Sigma]_{\vartheta,\vec{\mf{w}}} -E_C[\emptyset]&=\frac{1}{2}(\mf{b}+\mf{b}^{-1})^2[(\hat\varrho_{\mf{l}},\hat\varrho_{\mf{l}}) - (\hat\varrho_{\mf{g}},\hat\varrho_{\mf{g}})]+\frac{1}{2}(\vec{\mf{w}},\vec{\mf{w}}) ,\\\nonumber
      &=-\frac{1}{6}\left(N^3 - \sum_{a=1}^n N_a^3 - 3(\vec{\mf{w}},\vec{\mf{w}})\right).
 \end{align}
 In the second line we took the limit $\mf{b}\to 1$, and replaced the scalar product of the Weyl vectors -- denoted $\hat\varrho_{\mf{l}}$ and $\hat\varrho_{\mf{g}}$ for $\mf{l}$ and $\mf{g}=\mf{su}(N)$, respectively -- with 
 \begin{align}
     (\hat\varrho_{\mf{l}},\hat\varrho_{\mf{l}}) = \frac{1}{12}\sum_{a=1}^n(N_a^3-N_a),\qquad (\hat\varrho_{\mf{g}},\hat\varrho_{\mf{g}}) = \frac{1}{12}(N^3-N).
 \end{align}
 Turning off the monodromy parameters\footnote{In light of the compact LLM-type solutions found recently in \cite{Bomans:2023ouw} where the additional internal $U(1)$ symmetry is broken by the presence of scalar fields, which are interpreted as monodromy parameters, it may be possible to pin down a more precise relation between $E_C$ and defect anomalies by computing $\left< T_{\mu\nu}\right>$ if similar non-compact solutions allowing for $\mf{w}_a\neq 0$ can be constructed.} ($\mf{w}_a=0$) in \eq{Defect-SCE} we see the superficial relation
 \begin{align}
   E_C[\Sigma]_{\vartheta,\vec{0}} -E_C[\emptyset] = 4d_2|_{k_a\to1}~,
 \end{align}
 where on the right hand side we take all orbifold parameters $k_a\to1$ in \eq{Electrostatic-d2-2}.  

Since the expression for the defect SCE in terms of explicit defect Weyl anomalies is still unknown and $4d$ DCFTs have 23 possible parity even anomalies, we cannot definitively state that the defect SCE is determined solely by $d_2$.  We note, though, that a similar relation was found for co-dimension 4 Wilson surface defects: the defect SCE in that case was also related to the $2d$ DCFT equivalent of $d_2$.  Since $2d$ DSCFT preserving at least $\CN=(2,0)$ supersymmetry have only two independent Weyl anomalies\footnote{This was first proven for superconformal surface defects in $4d$ $\CN=2$ SCFTs in \cite{Bianchi:2019sxz}, and later, it was proven for 2d defects in the $6d$ $\CN=(2,0)$ theory in \cite{Drukker:2020atp}. }, which for the Wilson surface defect can be clearly distinguished from one another \cite{Jensen:2018rxu}, it was conjectured that $d_2$ alone fixed the defect SCE \cite{Chalabi:2020iie}. So, while it is not inconceivable that $d_2$ could appear in the defect SCE for co-dimension 2 defects, we leave establishing the precise relation for future work.

\subsubsection*{R-anomalies}

Ordinarily in $4d$ SCFTs, there are non-perturbative formulae that relate the A-type and B-type Weyl anomalies to `t Hooft anomalies for the superconformal $R$ symmetry \cite{Anselmi:1997am}.  In \cite{Wang:2021mdq}, it was conjectured that $a_\Sigma$ obeys the same relation to defect $R$-anomalies as a standalone theory\footnote{It was also conjectured that a B-type defect anomaly built out of the square of intrinsic Weyl tensor ($c_\Sigma|\bar{W}|^2$) obeys the usual relation \cite{Anselmi:1997am}
\begin{align}
    c_\Sigma =\frac{9k_{rrr}-5k_r}{32}. 
\end{align}
However, the basis used in \cite{Chalabi:2021jud} did not include $|\bar{W}|^2$.  From the Gauss-Codazzi and Ricci relations, $|\bar{W}|^2$ is related to several anomalies in the original basis (none of which include $d_2$).  So it is unclear at the this time, what observables can be used to compute $c_\Sigma$.  Though it is reasonable to expect that the defect limit of $\left< T_{\mu\nu}T_{\rho\sigma}\right>$ may be the appropriate correlator to compute $c_
\Sigma$, proving this is the subject of future work. }:
\begin{align}\label{eq:YW-a-Sigma}
    a_\Sigma = \frac{9k_{rrr}-3k_r}{32},
\end{align}
where $k_{rrr}$ and $k_r$ are the cubic and mixed $U(1)_r$ R-anomalies. Importantly for the defect theory written in $4d$ $\CN=1$ language, the superconformal $r_\Sigma$ symmetry is a linear combination of the Cartan generator of the ambient $SU(2)_R$ R-symmetry and the generator of normal bundle rotation $M_\varphi$ \cite{Wang:2021mdq}
\begin{align}
    r_{\Sigma} = \frac{2}{3}(2r_{6d} - M_\varphi).
\end{align}
It was further stated in \cite{Wang:2021mdq} that precisely for the types of defects holographically described by the electrostatic solutions considered above, in order to determine the $R$ and mixed anomaly we should use the counting formulae \cite{Bah:2019jts}
\begin{align}\label{eq:Bah-R-anomalies}
    k_{rrr} = \frac{2}{27}(\mf{n}_v-\mf{n}_h)+\frac{8}{9}\mf{n}_v,\qquad k_r = \frac{2}{3}(\mf{n}_v-\mf{n}_h),
\end{align}
where $\mf{n}_v$ is the number of $4d$ vector multiplets and $\mf{n}_h$ is the number hypermultiplets. In turn, both $\mf{n}_h$ and $\mf{n}_v$ are determined by the Young diagram data.

As we have pointed out around \eq{Electrostatic-a-3}, the defect A-type anomaly contains a contribution that is precisely of the form of the central charge $c_{\rm 4d}$ of $4d$ SCFTs engineered from irregularly punctured Riemann surface compactifications of $6d$ $\CN=(2,0)$ $A_{N-1}$ series SCFTs dual to electrostatic solutions of the type studied above.  Further, in \cite{Bah:2019jts,Bah:2022yjf}, a match was found between the holographic computation of $c_{\rm 4d}$ of the dual $4d$ SCFT and the large $N$ behavior of the central charge computed in the field theory using the R-anomalies and \eq{Bah-R-anomalies}.  However since we have found $a_\Sigma = c_{\rm 4d} + N^3/32$, it is clear that the na\"ive application of \eq{YW-a-Sigma} and \eq{Bah-R-anomalies} do not directly match.

\subsection{Future directions and open questions}\label{sec:Discussion-Future-directions}

The work that we have presented in this paper is only scratching the surface of $4d$ defects. While a full accounting of all of the defect Weyl anomalies of these systems through computing entropies, correlation functions, or other physical quantities is not currently possible, there are a number of questions opened up by our analysis that we will leave for future work.

\subsubsection*{Probe branes}

Even though we have access to the full $11d$ SUGRA bubbling geometry solution, it is useful to consider limit cases where we can instead appeal to a probe brane construction.  By finding $\kappa$-symmetric embeddings of probe M5-branes in an AdS$_7\times\mathds{S}^4$ background wrapping AdS$_5\subset$ AdS$_7$ and an $\mathds{S}^1$ living either in the internal $\mathds{S}^4$ or in the AdS$_7$, we expect to be able to holographically study defects engineered by Young diagrams associated to totally symmetric or totally antisymmetric representations of $\mf{su}(N)$ similar to the co-dimension 4 Wilson surface defects from M2 and M5 probe branes \cite{Mori:2014tca, Agarwal:2018tso, Rodgers:2018mvq}.  One advantage of studying these defect systems using probe brane holography is that we will have clearer access to the study of defect RG flows, which will provide holographic tests of the defect $a_\Sigma$-theorem in a strongly coupled theory, a means to study defect phase transitions, and a setting to test the monotonicity of the defect sphere EE along an RG flow \cite{Rodgers:2018mvq}.  Further taking inspiration from AdS$_5$ holography \cite{Karch:2015vra,Karch:2015kfa,Robinson:2017sup}, if one was able to construct a $\kappa$-symmetric probe M5 brane embedding in global AdS$_7$, say with an $\mathds{S}^1\times\mathds{S}^5$ boundary, one could try to compare to recent results in type IIB probe brane holography and supersymmetric localization in $3d/5d$ systems on a sphere \cite{Gutperle:2020rty,Santilli:2023fuh}.  These questions are currently being investigated in work currently in progress.

\subsubsection*{Dimensional reduction}

By (partial) topologically twisted dimensional reduction on a Riemann surface or a 3-manifold, $6d$ SCFTs can be used to engineer large classes of $4d$ \cite{Gaiotto:2009fs, Gaiotto:2015usa} and $3d$ \cite{Terashima:2011qi,Dimofte:2011ju} theories.  Further, we can enrich the algorithm to determine the lower dimensional theory by starting from a $6d$ theory deformed by their natural co-dimension 2 and 4 defects to end up with a dimensionally reduced theory possibly with defects \cite{Alday:2009fs,Ito:2016fpl,Gang:2015wya}.  As we have seen in the computation of the A-type anomaly for co-dimension 2 defects in the $6d$ $\CN=(2,0)$ $A_{N-1}$ series SCFTs, there is a connection to the central charge of a $4d$ SCFT engineered on a Riemann surface with regular punctures, at least in the large $N$ limit. It is natural, then, to wonder how the rest of the data contained in the other 22 parity even defect Weyl anomalies can be used to characterize the lower dimensional theory, or whether the remaining unknown defect Weyl anomalies are vanishing or fixed by $a_\Sigma$ and $d_2$.  For BPS Wilson surfaces in $6d$ preserving at least $2d$ $\CN=(2,0)$ supersymmetry, the defect supersymmetry imposes non-trivial relations among the B-type defect Weyl anomalies \cite{Drukker:2020atp}, but as of yet, there is no known relation imposed by $4d$ $\CN=2$ defect supersymmetry.

A special case of dimensional reduction of the $6d$ $\CN=(2,0)$ $A_{N-1}$ theory is taking the Riemann surface to be $\mathds{T}^2$, which reduces to $4d$ $\CN=4$ $SU(N)$ super Yang-Mills theory.  The co-dimension 2 defects labelled by $\vartheta:\mf{sl}(2)\to \mf{su}(N)$ in the parent theory that we have holographically studied above wrapped on $\mathds{T}^2$ reduce to Gukov-Witten type defects.  In the absence of complex structure deformations on $\mathds{T}^2$, all of the defect Weyl anomalies are equal to one another and are $\propto N^2 -\sum_a N_a^2$ \cite{Gomis:2007fi,Drukker:2008wr, Gentle:2015ruo,Jensen:2018rxu}, which is closer in appearance to $d_2$ in \eq{Electrostatic-d2-2} than $a_\Sigma$ in \eq{Electrostatic-a-3}.  However, an exact relation to determine the anomalies of the Gukov-Witten defect from the higher dimensional defect anomalies is as of yet unknown.

\subsubsection*{Defect Weyl anomalies and `t Hooft anomalies}

As we saw in the attempt to match the any of the holographic results for $a_\Sigma$ or $d_2$ to large $N$ field theory computations, there are points of tension that should be resolved. One of the biggest issues, though, is that the putative relation between defect `t Hooft anomalies and defect Weyl anomalies seemed to disagree with the holographic results.  While it remains a possibility that the issue stems from the holographic side of the story, there is an open question on the field theory side that must be addressed as well.  Namely, the formulae conjectured in \cite{Wang:2021mdq} only relate two of the twenty-three parity even defect Weyl anomalies to the defect R-anomalies for co-dimension $\geq 2$ $4d$ defects.  That is, only the $\overline{E}_4$ and $|\overline{W}|^2$ structures in the defect anomaly have been supersymmetrized.  A similar supersymmetrization of the defect Weyl anomaly for $2d$ defects limited to be sensitive only to the intrinsic geometry of the defect submanifold was carried out in \cite{Wang:2020xkc}.  This naturally leads one to wonder if it is possible to supersymmetrize the full defect Weyl anomaly including the anomalies containing the second fundamental form and normal bundle curvature in order to arrive at a complete set of non-perturbative formulae for defect Weyl anomalies.

\section*{Acknowledgments}The authors would like to thank Pieter Bomans,  Michael Gutperle, and Andy O'Bannon for useful discussions throughout this work.  We would also like to thank Pieter Bomans 
and Michael Gutperle 
for comments on the draft.
JE would also like to thank the EIC Theory Institute at BNL for partial financial support and warm hospitality while this work was being completed.  The work of PC is supported by a Mayflower studentship from the University of Southampton. The work of BR is supported by the INFN. The work of BS is supported in part by the STFC consolidated grant ST/T000775/1.  This material is based upon work supported by the U.S. Department of Energy, Office of Science, Office of High Energy Physics under Award Number DE-SC0024557.

Disclaimer:  ``This report was prepared as an account of work sponsored by an agency of the United States Government.  Neither the United States Government nor any agency thereof, nor any of their employees, makes any warranty, express or implied, or assumes any legal liability or responsibility for the accuracy, completeness, or usefulness of any information, apparatus, product, or process disclosed, or represents that its use would not infringe privately owned rights.  Reference herein to any specific commercial product, process, or service by trade name, trademark, manufacturer, or otherwise does not necessarily constitute or imply its endorsement, recommendation, or favoring by the United States Government or any agency thereof.  The views and opinions of authors expressed herein do not necessarily state or reflect those of the United States Government or any agency thereof.''

\appendix
\section{Fefferman-Graham coordinates}\label{app:FG}
The starting point for computing holographic quantities associated with the two-charge solutions and electrostatic solutions is finding the asymptotic transformation which maps their respective metrics into FG gauge.  In this appendix, we will first derive the transformations of \eq{Two-charge-metric} and find the asymptotic expressions for the metric functions in FG gauge.  We will also derive the transformation of \eq{LLM-defect-metric} into FG gauge. In this process, we will find necessary conditions on the mixing of two of the angular coordinates that allow for the metrics to be put into FG form.  The interpretation of this mixing of angular coordinates is interpreted in the field theory language as an identification of the defect superconformal R-symmetry.
\subsection{Two-charge solutions}\label{app:Two-charge-FG}

In this subsection, we will focus on putting the two-charge solutions in FG gauge.  The explicit forms of the metric functions in \eq{Two-charge-metric} are as follows:
\begin{subequations}\allowdisplaybreaks\label{eq:Two-charge-fs}
    \begin{align}\allowdisplaybreaks
        \hat{f}_{\rm AdS}^2=&\ \kappa^{2/3}\left[\frac{c_{\zeta }^2 \left(q_1+y^2\right) \left(q_2-q_2 c_{2 \psi }+2 y^2\right)}{2 y}+y \left(q_2+y^2\right) s_{\zeta }^2\right]^{1/3}~,\\
        \hat{f}_y^2=&\ \kappa^{2/3}\frac{\hat{f}_{\rm AdS}^2  y}{4 \left(q_1+y^2\right) \left(q_2+y^2\right)-4 y^3}~,\\
        \hat{f}_z^2=&\ \kappa^{2/3}\Biggr[\frac{c_{\zeta }^2 \left(c_{2 \psi } \left( \left(a_2+1\right){}^2 q_2 y+ a_2^2 y^3- q_2 \left(q_1+y^2\right)\right)+ \left(a_2+1\right){}^2 q_2 y+ \left(a_2^2-2\right) y^3\right)}{2y \hat{f}_{\rm AdS}^4}\nonumber\\
        &+\frac{ s_{\zeta }^2 \left(y \left( \left(a_1^2-1\right) y+ \left(q_2+y^2\right)\right)+ \left(a_1+1\right){}^2 q_1\right)}{\hat{f}_{\rm AdS}^4}+\frac{c_{\zeta }^2 \left(q_1+y^2\right) \left(q_2+2 y^2\right)}{2\hat{f}_{\rm AdS}^4y}\Biggr]~,\\
        \hat{f}_{\phi_1}^2=&\ \kappa^{2/3}\frac{\left(q_1+y^2\right) s_{\zeta }^2}{4 \hat{f}_{\rm AdS}^4}~,\\
        \hat{f}_{\phi_2}^2=&\ \kappa^{2/3}\frac{c_{\psi }^2 c_{\zeta }^2 \left(q_2+y^2\right)}{4 \hat{f}_{\rm AdS}^4}~,\\
        \hat{f}_{z\phi_1}^2=&\ \kappa^{2/3}\frac{s_{\zeta }^2 \left(a_1 q_1+a_1 y^2+q_1\right)}{\hat{f}_{\rm AdS}^4}~,\\
        \hat{f}_{z\phi_2}^2=&\ \kappa^{2/3}\frac{ c_{\psi }^2 c_{\zeta }^2 \left(a_2 q_2+a_2 y^2+q_2\right)}{ \hat{f}_{\rm AdS}^4}~,\\
        \hat{f}_\psi^2=&\ \kappa^{2/3}\frac{c_{\zeta }^2 \left(q_2-q_2 c_{2 \psi }+2 y^2\right)}{8\hat{f}_{\rm AdS}^4}~,\\
        \hat{f}_\zeta^2=&\ \kappa^{2/3}\frac{q_1 c_{2 \zeta }+2 q_2 c_{\psi }^2 s_{\zeta }^2+q_1+2 y^2}{8 \hat{f}_{\rm AdS}^4}~,\\
        \hat{f}_{\psi\zeta}^2=&\ \kappa^{2/3}\frac{ q_2 c_{\psi } c_{\zeta } s_{\psi } s_{\zeta }}{2 \hat{f}_{\rm AdS}^4}~,
    \end{align}
\end{subequations}
where we denote $\kappa=\hat{g}^3N\ell_{\text{P}}^3/2$.

We seek an asymptotic map from $\{y,\,\psi,\,\zeta\}$ to the FG coordinates $\{u, \aleph, \theta\}$ in the  large-$y$/small-$u$ regime. By solving
\begin{align}
\hat{f}_y^2 dy^2 +\hat{f}^2_\psi d\psi^2+\hat{f}^2_\zeta d\zeta^2 +  \hat{f}^2_{\psi\zeta}d\psi d\zeta = \frac{L^2}{u^2}du^2+\frac{L^2}{4}\left(c_\theta^2\hat{\alpha}_\aleph d\aleph^2 +\hat{\alpha}_\theta d\theta^2 + \hat{\alpha}_{\theta\aleph}d\theta d\aleph\right)   
\end{align}
order by order in $u$, we find that the appropriate asymptotic map is
\begin{align}\label{eq:Two-charge-FG-transformation}
    \begin{split}
        y&=\frac{1}{u^2}+\frac{1}{2}+\frac{\left(2 q_1-q_2\right) c_{2
        \theta }-2 q_2 c_{2 \aleph } c_{\theta }^2-10 q_1-9 q_2+3}{48} u^2+\ldots,\\
        \psi&=\aleph+\frac{q_2  s_{2 \aleph }}{24}u^4+\ldots,\\
        \zeta&=\theta-\frac{s_{2 \theta } \left(q_1-q_2 c_{\aleph }^2\right)}{24}u^4+\ldots~,
    \end{split}
\end{align}
 where we have suppressed higher orders in $u$ due to their cumbersome expressions.  To complete this map, we need to identify $\kappa=L^3$, where $L$ denotes the radius of the asymptotic AdS$_7$ spacetime. 

Mapping all of the other metric functions in \eq{Two-charge-metric}, we find the FG form of the metric to be
\begin{align}
\begin{split}\label{eq:Two-charge-FG-metric}
    ds^2_{\rm FG} &= \frac{L^2}{u^2}(du^2 + \hat{\alpha}_{\rm AdS}ds_{\rm AdS_5}^2+\hat{\alpha}_z dz^2)+L^2s_\theta^2\hat{\alpha}_{z\varphi_1}dzd\varphi_1 + L^2c_\aleph^2c_\theta^2\hat{\alpha}_{z\varphi_2}dzd\varphi_2\\
    &+\frac{L^2}{4}(\hat{\alpha}_\theta d\theta^2 +s_\theta^2\hat{\alpha}_{\varphi_1}d\varphi_1^2 + c_\theta^2(\hat{\alpha}_\aleph d\aleph^2+c_\aleph^2\hat{\alpha}_{\varphi_2}d\varphi_2^2) + \hat{\alpha}_{\theta\aleph}d\theta d\aleph),
\end{split}
\end{align}
where we have transformed the angular coordinates using
\begin{align}\label{eq:Two-charge-new-phi}
    \phi_I = \varphi_I -2a_I z.
\end{align}
Note that since $\phi_I$ and $z$ are all $2\pi$-periodic and $a_I\in\mathbb{Z}/2$, the new angular coordinates $\varphi_I$ are also $2\pi$-periodic. The metric functions have the asymptotic behavior
\begin{subequations}\label{eq:Two-charge-FG-alphas}\allowdisplaybreaks
    \begin{align}\allowdisplaybreaks
        \hat{\alpha}_{\rm AdS} &=1+\frac{u^2}{2} +\frac{3-2q_1+3q_2-10q_2 c_{2\aleph}c_\theta^2+5(2q_1-q_2)c_{2\theta}}{48}u^4+\ldots,\\
        \hat{\alpha}_z &=1-\frac{u^2}{2}+\frac{3-2q_1+3q_2-10q_2c_{2\aleph}c_\theta^2+5(2q_1-q_2)c_{2\theta}}{48}u^4+\ldots,\\
        \hat{\alpha}_{\varphi_1}&=1+\frac{10q_2c_{2\aleph}c_{\theta}^2+5(q_2-2q_1)c_{2\theta}+14q_1-11q_2}{24}u^4+\ldots,\\
        \hat{\alpha}_{\varphi_2}&=1+\frac{10q_2c_{2\aleph}c_{\theta}^2+5(q_2-2q_1)c_{2\theta}-6q_1+9q_2}{24}u^4+\ldots,\\
        \hat{\alpha}_{z\varphi_1}&=q_1u^4-q_1u^6+\ldots,\\
        \hat{\alpha}_{z\varphi_2}&=q_2u^4-q_2u^6+\ldots,\\
        \hat{\alpha}_{\theta}&=1+\frac{5q_2c_{2\aleph}+2q_1-3q_2}{12}u^4+\ldots,\\
        \hat{\alpha}_{\aleph}&=1+\frac{5(q_2-2q_1)c_{2\theta}-10q_2c_{2\aleph}s_{\theta}^2-6q_1-q_2}{24}u^4+\ldots,\\
        \hat{\alpha}_{\theta\aleph}&=\frac{5q_2 s_{2\theta}s_{2\aleph}}{12}u^4+\ldots~.
    \end{align}
\end{subequations}

If we had not transformed to $\varphi_I$, we would not have been able to put the metric in FG form.  We can see this in the original $\phi_I$ coordinates, where $\hat{\a}_{z\phi_I}$ has an $O(1)$ term which is proportional to $a_I$. FG gauge requires $\hat{\a}_{z\phi_I}\sim u^4$, which would mean setting $a_I=0$. However, the values of the $a_I$'s are set by regularity, i.e. 
\begin{align}
    a_I = -\frac{q_I}{q_I+y_+^2}~,
\end{align}
and so, we cannot simply tune them to zero without also setting the corresponding $q_I=0$, which lands us on the pure AdS$_7\times \mathds{S}^4$ solution.

\subsection{Electrostatic solutions}\label{app:Electrostatic-FG}

We now turn to deriving the FG form of the metric for the electrostatic solutions.  Finding the asymptotic expansions of the metric factors in \eq{LLM-defect-metric} requires explicit expressions for $\dot{V}$, $\ddot{V}$, $\dot{V}^\prime$, $V^{\prime\prime}$, and $\sigma$. We can compute the indefinite integral in $V$ for a trial line charge distribution $\varpi_a(\eta) = p_{1+a} \eta + \delta_{1+a}$,
\begin{align}
    \hspace{-0.45cm} -\frac{1}{2}\int d\eta^\prime G(r,\eta,\eta^\prime) \varpi_a(\eta^\prime)  =&\ \frac{p_{1+a}}{2}\bigg(\sqrt{r^2+(\eta+ \eta^\prime)^2}-\sqrt{r^2+(\eta- \eta^\prime)^2}\\\nonumber
     &- \eta \tanh^{-1}\Big(\frac{\eta+\eta^\prime}{\sqrt{r^2+(\eta+ \eta^\prime)^2}}\Big)+\eta \tanh^{-1}\Big(\frac{\eta-\eta^\prime}{\sqrt{r^2+(\eta- \eta^\prime)^2}}\Big)\bigg)\\\nonumber
    &+\frac{\delta_{1+a}}{2}\left(\tanh^{-1}\Big(\frac{\eta+\eta^\prime}{\sqrt{r^2+(\eta+ \eta^\prime)^2}}\Big) + \tanh^{-1}\Big(\frac{\eta-\eta^\prime}{\sqrt{r^2+(\eta- \eta^\prime)^2}}\Big)\right)~,
\end{align}
and then build up the full potential by summing over the intervals.  Clearly, evaluating the result above in the $\eta^\prime \to \infty$ region leads to linear and logarithmic divergences. However, when evaluating derivatives of the right-hand side above, these divergences are eliminated, and only derivatives of $V$ appear in all of the computations carried out below and in the main body of the text.

The asymptotically AdS$_7\times \mathds{S}^4$ region corresponds to the limits $r,\,\eta\to\infty$.  In order to facilitate the expansion of the derivatives of the electrostatic potential in this region, we redefine $r = \varrho c_\omega$ and $\eta = \varrho s_\omega$, with $\omega \in [0,\pi/2]$, so that
\begin{align}
    f_3^2(dr^2 +d\eta^2) \to f_\varrho^2d\varrho^2+f_\omega^2d\omega^2,
\end{align}
with $f_\varrho^2 = f_3^2$ and $f_\omega^2 = f_3^2 \varrho^{2}$. The AdS$_7\times\mathds{S}^4$ region now lies in the $\varrho\to\infty$ limit. We can compute the asymptotic expansions of the derivatives of the electrostatic potential in this region in terms of its moments as follows,
\begin{subequations}\label{eq:Electrostatic-asymptotic-V}\allowdisplaybreaks
\begin{align}\allowdisplaybreaks
    \begin{split}
        \dot{V} &=\varrho  s_{\omega }+m_1 s_{\omega }-\frac{m_3 c_{\omega }^2 s_{\omega }}{2 \varrho ^2}+\frac{m_5 \left(7 c_{2 \omega }-1\right)
   c_{\omega }^2 s_{\omega }}{16 \varrho ^4}+\ldots~,
   \end{split}\\
   \begin{split}
   \ddot{V}&=-m_1 c_{\omega }^2 s_{\omega }+\frac{m_3 \left(5 c_{2 \omega }+1\right) c_{\omega }^2 s_{\omega }}{4 \varrho ^2}-\frac{m_5
   \left(28 c_{2 \omega }+63 c_{4 \omega }+29\right) c_{\omega }^2 s_{\omega }}{64 \varrho
   ^4}+\ldots~,
    \end{split}\\
    \begin{split}
    \dot{V}^\prime &=
    1+\frac{m_1 c_{\omega }^2}{\varrho }+\frac{m_3 \left(3-5 c_{2 \omega }\right) c_{\omega }^2}{4 \varrho ^3}+\frac{3 m_5 \left(21 c_{4 \omega }-28
   c_{2 \omega }+15\right) c_{\omega }^2}{64 \varrho ^5}+\ldots~,
    \end{split}\\
    \begin{split}
    V^{\prime\prime}&=\frac{m_1 s_{\omega }}{\varrho ^2}-\frac{m_3 \left(5 c_{2 \omega }+1\right) s_{\omega }}{4 \varrho ^4}+\frac{m_5 \left(28 c_{2
   \omega }+63 c_{4 \omega }+29\right) s_{\omega }}{64 \varrho ^6}+\ldots~.
    \end{split}
\end{align}
From these expressions, we can also find the asymptotic behavior of $\sigma$ in terms of the moments of the electrostatic potential to be
\begin{align}
       \sigma= 1+\frac{2 m_1}{\varrho
   }-\frac{m_1^2 \left(c_{2 \omega }-3\right)}{2 \varrho ^2}+\frac{m_3 \left(1-3 c_{2 \omega }\right)}{2
   \varrho ^3}
   +\frac{m_3 m_1 \left(1-12 c_{2 \omega
   }+3 c_{4 \omega }\right)}{8 \varrho ^4}+\ldots~.
\end{align}
\end{subequations}
Together, these expansions can be inserted into the definitions of the metric functions in \eq{LLM-defect-metric} to give 
\begin{subequations}\allowdisplaybreaks
\begin{align}\allowdisplaybreaks
  \hspace{-0.5cm} \frac{(2m_1)^{1/3}}{\k_{11}^{2/3}} f_{\rm AdS}^2 &= 4 \varrho +4 m_1+\frac{5 m_3 c_{2\omega }+4 m_1^3 s_{\omega }^2+m_3}{3 m_1 \varrho }+\frac{4  \left(m_3-m_1^3\right) s_{\omega }^2}{3 \varrho ^2}+\ldots~,\\
\hspace{-0.5cm}\frac{(2m_1)^{1/3}}{s_\omega^2\k_{11}^{2/3}}f_{\mathds{S}^2}^2&= 2 m_1-\frac{ \left(1+5 c_{2 \omega }\right)m_3+4 s_{\omega }^2m_1^3 }{3 \varrho ^2}+\frac{8
   m_1 \left(m_1^3-m_3\right) s_{\omega }^2}{3 \varrho ^3}+\ldots~,\\
\hspace{-0.5cm}\frac{(2m_1)^{1/3}}{\k_{11}^{2/3}}f_\varrho^2&=\frac{2 m_1}{\varrho ^2}-\frac{(1+5c_{2\omega})m_3 - 2s_\omega^2m_1^3}{3 \varrho^4}+\frac{4 m_1 \left(m_3-m_1^3\right) s_{\omega }^2}{3 \varrho ^5}+\ldots~,\\
\hspace{-0.5cm}\frac{(2m_1)^{1/3}}{\k_{11}^{2/3}}f_\beta^2&= 4 \varrho +m_1 \left(c_{2 \omega }-3\right)+\frac{(1+5 c_{2 \omega })m_3 +4 m_1^3 s_{\omega}^2}{3 m_1 \varrho }+\ldots~,\\
\hspace{-0.5cm}\frac{(2m_1)^{1/3}}{\k_{11}^{2/3}}f_\chi^2&= 4 \varrho +4 m_1 c_{2 \omega }+\frac{5 m_3 c_{2 \omega }+4 m_1^3 s_{\omega }^2+m_3}{3 m_1
   \varrho }+\ldots~,\\
\hspace{-0.5cm}\frac{(2m_1)^{1/3}}{\k_{11}^{2/3}}f_{\beta\chi}^2&= 8 \varrho -8 m_1 s_{\omega }^2+\frac{2 \left(5 m_3 c_{2 \omega }+4 m_1^3
   s_{\omega }^2+m_3\right)}{3 m_1 \varrho }+\ldots~.
\end{align}
\end{subequations}

We again look for an asymptotic map to a set of coordinates $\{u,\theta\}$ in terms of which the metric is in FG form. By taking $\varrho = \varrho(u,\theta)$ and $\omega = \omega(u,\theta)$, and expanding in small $u$ to solve
\begin{align}
    f_\varrho^2d\varrho^2 + f_\omega^2d\omega^2 = \frac{L^2}{u^2}du^2+\frac{L^2}{4}\alpha_\theta d\theta^2
\end{align}
order by order, we find
\begin{subequations}\label{eq:Electrostatic-FG-transformation}
    \begin{align}
        \rho&=\frac{2 m_1}{u^2}+\frac{2 m_1^3 c_{\theta }^2+m_3 \left(5 c_{2 \theta }-1\right)}{48 m_1^2}u^2+\frac{\left(m_3-m_1^3\right)c_{\theta }^2}{36 m_1^2}u^4+\ldots~,\\
        \omega&=\theta+\frac{\pi }{2} -\frac{\left(m_1^3+5 m_3\right) s_{2 \theta }}{96 m_1^3}u^4 +\frac{\left(m_1^3-m_3\right) s_{2 \theta }}{216 m_1^3}u^6+\ldots~.
\end{align}
\end{subequations}
The asymptotic expansions of $f_\chi^2$, $f_\beta^2$, and $f_{\beta\chi}^2$ under the above transformation reveal an ambiguity as to which of the angular coordinates should be identified as parametrizing the external $\mathds{S}^1\subset$ AdS$_7$ and which as parametrizing the internal $\mathds{S}^1\subset\mathds{S}^4$ upon mapping to FG gauge. That is, both are characterized by $1/u^2$ divergences at small-$u$, so that the resulting asymptotic metric is not in FG gauge. To resolve this issue, we introduce
\begin{align}\label{eq:Electrostatic-z-varphi-transformation}
    \chi = (1+\CC_z)z + a_\varphi \varphi,\qquad \beta = -\CC_zz + b_\varphi \varphi,
\end{align}
where $\CC_z\in \mathbb{Z}$ and $a_\varphi$ and $b_\varphi$ are arbitrary constants. Note that this transformation parallels the one taken in \cite{Bah:2022yjf}, where $\CC_z = 1/\CC$ is fixed by the ratio of four-form flux through two 4-cycles, which in turn fixes the mixing parameter between the $U(1)$ symmetries leading to $U(1)_r$ symmetry $\pd_\chi = \pd_z + \frac{1}{\CC}\pd_\varphi$ in the field theory.  Here we are following the conventions of \cite{Gutperle:2023yrd} where the corresponding $\CC$ is negative. 
We then find that the metric functions for the transformed coordinates display the following asymptotic behavior,
\begin{subequations}
\begin{align}
    \frac{f_\varphi^2}{L^2} &= \frac{(a_\varphi+b_\varphi)^2}{u^2}-\frac{1}{8}\left((2a_\varphi+b_\varphi)^2c_{2\theta}+b_\varphi(4a_\varphi+3b_\varphi)\right)+\ldots~,\\
    \frac{f_{z\varphi}^2}{L^2} &=\frac{2(a_\varphi+b_\varphi)}{u^2}+\frac{1}{4}(2a_\varphi\mc{C}_z+b_\varphi(\mc{C}_z-2)-(2a_\varphi+b_\varphi)(\mc{C}_z+2)c_{2\theta})+\ldots~,\\
    \frac{f_{z}^2}{L^2} &= \frac{1}{u^2} +\frac{1}{8}(\mc{C}_z(\mc{C}_z+4)-(\mc{C}_z+2)^2c_{2\theta})+\ldots~,
\end{align}
\end{subequations}
where we introduced the AdS$_7$ radius $L=(16m_1\kappa_{11})^{1/3}$. Setting $a_\varphi= -b_\varphi =-1$ removes the $1/u^2$ divergences in the asymptotic expansions of $f_\varphi^2$ and $f^2_{z\varphi}$. In particular, $f_\varphi^2=L^2s_\theta^2/4+\ldots$. This identifies the $\varphi$-circle as the internal $\mathds{S}^1\subset\mathds{S}^4$. Furthermore, $f_z^2 = L^2/u^2+\ldots$, as required for the external $\mathds{S}^1\subset$ AdS$_7$. The final requirement to achieve an FG parametrization is that $f_{z\varphi}^2 \sim O(u^2)$.  Eliminating the  $u^0$ behavior of $f_{z\varphi}^2$ fixes $\mc{C}_z\equiv-2$. Recalling the role of $\CC_z$, we see that the defect superconformal R-symmetry is $\pd_\chi = \pd_z - 2\pd_\varphi$. 

Having identified the correct combination of angular variables, we can at once express the metric in FG gauge as
\begin{align}\begin{split}\label{eq:Electrostatic-FG-metric}
    ds_{\rm FG}^2 &= \frac{L^2}{u^2}(du^2 + \alpha_{\rm AdS}ds_{\text{AdS}_5}^2 +\alpha_z dz^2) +L^2s_\theta^2\alpha_{z\varphi}dz d\varphi\\
    &\quad+\frac{L^2}{4}\big(s_\theta^2\alpha_\varphi  d\varphi^2 +c_\theta^2 \alpha_{\mathds{S}^2}d\Omega_2^2 +\alpha_\theta d\theta^2 \big)~,
\end{split}\end{align}
where the metric functions have asymptotic behavior
\begin{subequations}\allowdisplaybreaks
\begin{align}\allowdisplaybreaks
    \alpha_{\text{AdS}}&=1+\frac{u^2}{2}+\frac{1}{96} \left(10 c_{\theta }^2+\frac{m_3 \left(1-5 c_{2 \theta }\right)}{m_1^3}\right)u^4+\frac{\left(m_3-m_1^3\right)  c_{\theta }^2}{18 m_1^3}u^6\ldots~,\\
    \alpha_{z}&=1-\frac{u^2}{2}+\frac{1}{96} \left(10 c_{\theta }^2+\frac{m_3 \left(1-5 c_{2 \theta }\right)}{m_1^3}\right)u^4+\frac{\left(m_3-m_1^3\right) \left(5 c_{2 \theta }-13\right)}{72 m_1^3}u^6+\ldots~,\\
    \alpha_{\varphi}&=1+\frac{\left(m_3-m_1^3\right) \left(5 c_{2 \theta }-7\right)}{48 m_1^3}u^4+\frac{\left(m_1^3-m_3\right)\left(10 c_{2 \theta }-17\right)}{108 m_1^3}u^6+\ldots~,\\
    \alpha_{\mathds{S}^2}&=1+\frac{\left(m_3-m_1^3\right) \left(5 c_{2 \theta }+3\right)}{48 m_1^3}u^4+\frac{\left(m_1^3-m_3\right) \left(5 c_{2 \theta }+4\right)}{54 m_1^3}u^6+\ldots~,\\
    \alpha_{z\varphi}&=\frac{m_1^3-m_3}{4 m_1^3}u^4-\frac{m_1^3-m_3}{4m_1^3}u^6+\ldots~,\\
    \alpha_\theta  &= 1 +\frac{m_1^3-m_3}{24m_1^3} u^4+\frac{\left(m_3-m_1^3\right) \left(5 c_{2 \theta }+9\right)}{216 m_1^3} u^6+\ldots~.
\end{align}
\end{subequations}
Note that, upon being evaluated on the single kink electrostatic profile in \eq{singlekink}, the asymptotic metric above recovers the $q_2=0$ instance of \eq{Two-charge-FG-metric}; in particular, the coordinate $\varphi$ maps over to $\varphi_1$, while $\aleph$ and $\varphi_2$ correspond to, respectively, the polar and azimuthal angles on the asymptotic internal $\mathds{S}^2\subset\mathds{S}^4$ in the electrostatic description.

\section{Regulating the on-shell action }\label{app:On-shell-regulation}

In this appendix, we collect some of the details of the regulating scheme for the computation of the on-shell $11d$ supergravity action evaluated on the two-charge solutions.  Below we compute the vacuum AdS$_7\times\mathds{S}^4$ on-shell action, which we use in the background subtraction scheme.  This value also provides a good diagnostic for the known limiting case, $q_I=a_I=0$ for the two charge solutions, that recovers the vacuum geometry.  We also briefly discuss computing the renormalized volume of the AdS$_5$ submanifold of the $11d$ spacetime.

\subsection{\texorpdfstring{AdS$_7\times \mathds{S}^4$}{AdS7 x S4}}\label{app:AdS7-on-shell-action}

In this subsection, we compute the on-shell action for the vacuum AdS$_7\times\mathds{S}^4$ geometry that we use in our background subtraction scheme.   The data which specifies this solution to the bosonic theory in \eq{SUGRA-action-11d} is the metric
\begin{align}
ds_{11}^2 = L^2 \left(dx^2 + \cosh^2(x) ds_{\rm AdS_5}^2 + \sinh^2(x) dz^2 \right) + \frac{L^2}{4} d\Omega_4^2~,
\end{align}
which is an AdS$_5$ slicing of AdS$_7$ with $x\in[0,\infty)$, and the four-form flux and its Hodge dual
\begin{subequations}
    \begin{align}\label{eq:Vacuum-on-shell-F4}
    F_4&= -\frac{3L^3}{8} \Upsilon_{\tiny{\mathds{S}^4}}~, \\\label{eq:Vacuum-on-shell-star-F4}
    \star_{11}F_4&= 6 L^6 \cosh^5(x) \sinh(x) dx \wedge dz \wedge \Upsilon_{\rm AdS_5}~.
    \end{align}
\end{subequations}

Since we are working with the AdS$_7\times\mathds{S}^4$ vacuum, the transformation to FG gauge is simply
\begin{align}
x = -\ln(u/2)~,
\end{align}
where the FG radial coordinate is valued  $u\in[0,2]$.  In FG gauge, the metric takes the form
\begin{align}
ds_{11}^2 = \frac{L^2}{u^2} \left(du^2 + \left(1 + \frac{u^2}{2} + \frac{u^4}{16} \right) ds_{\rm AdS_5}^2 + \left(1 - \frac{u^2}{2} + \frac{u^4}{16}\right) dz^2 \right) + \frac{L^2}{4} d\Omega_4^2~.
\end{align}
The four-form flux has no functional dependence on $x$ and is unchanged in transforming to FG gauge, while the seven-form flux becomes
    \begin{align}\label{eq:Vacuum-on-shell-star-F4-FG}
    \star_{11}F_4&= 6 L^6 \left(  \frac{1}{u^7} + \frac{1}{u^5} + \frac{5}{16 u^3} - \frac{5 u}{256} - \frac{u^3}{256} - \frac{u^5}{4096}
    \right) du \wedge dz \wedge \Upsilon_{\rm AdS_5}+\ldots~.
    \end{align}

With the asymptotics of the metric and fluxes in hand, we can easily compute the on-shell action.  Note that the GHY term for the vacuum AdS$_7\times\mathds{S}^4$ solution is trivially identical to the expression found in \eq{Two-charge-on-shell-action-GHY}, and so we will not reproduce it here.  The bulk action is then computed from the $F_4\wedge\star F_4$ term, which after inserting eqs.~(\ref{eq:Vacuum-on-shell-F4}) and (\ref{eq:Vacuum-on-shell-star-F4-FG}), introducing a radial cutoff at $u=\eps_u\ll 1$, and integrating over the AdS$_7\times\mathds{S}^4$ geometry gives
\begin{align}
    S_{\rm{OS},\rm{bulk}}^{\rm{(vac)}}
    &= -\frac{L^{9} \pi^2}{8 G_{N}^{(11)}} \operatorname{vol}(\text{AdS}_5)  
    \left( \frac{1}{3 \epsilon_u^6} + \frac{1}{2 \epsilon_u^4} + \frac{5}{16 \epsilon_u^2}- \frac{11}{48} 
    \right)+ \ldots ~.
\end{align}
Combining with the GHY term, we find 
\begin{align}\label{eq:Vacuum-on-shell-action}
    S_{\rm{OS}}^{\rm{(vac)}}=-\frac{\pi^2 L^9}{8 G_{N}^{(11)}}\operatorname{vol}(\text{AdS}_5)\left(\frac{1}{3 \epsilon_u^6}+\frac{5}{3\epsilon_u^5}+\frac{1}{2 \epsilon_u^4}+\frac{1}{\epsilon_u^3}+\frac{5}{16 \epsilon_u^2}+\frac{5}{48\epsilon_u}-\frac{11}{48}\right)+\ldots~.
\end{align}

Finally, we note that since $d\star F_4=0$ we can introduce a gauge potential $C_6$ so that $d C_6 = \star_{11}F_4$.  We can then perform the bulk integral of $F_4\wedge C_6$ over the radial cutoff slice at $u=\eps_u$ with the pullback of the six-form potential being given by
\begin{align}
C_6\biggr|_{u=\epsilon_u} = 3 L^6 \left(  \frac{1}{3 \epsilon_u^6} + \frac{1}{2 \epsilon_u^4} + \frac{5}{16 \epsilon_u^2} - \frac{11}{48} + \frac{5 \eps_u^2}{256} + \frac{\eps_u^4}{512} + \frac{\eps_u^6}{12288}  \right) dz \wedge \Upsilon_{\rm AdS_5}.
\end{align}
Crucially, we have used the residual gauge freedom to fix the six-form potential to be regular at the origin of AdS$_7$, i.e. we pick a gauge such that $C_6\big|_{u=2} = 0$. In this gauge, the on-shell action computed using $C_6$ gives the same result as above.

\subsection{\texorpdfstring{Renormalized AdS$_5$ volume}{Renormalized AdS5 volume}}\label{app:AdS5-volume}

Even after accounting for the divergences coming from the asymptotically AdS$_7$ part of the geometry via background subtraction, we are still left to deal with the volume of the AdS$_5$ factor in the on-shell action.  In order to regularize the remaining polynomial divergences and read off the universal log-divergent part of the on-shell action, we will simply treat the intrinsic parts of the AdS$_5$ geometry using standard counterterms in holographic renormalization and neglecting any divergences associated with the embedding.  This renormalization scheme is admittedly simplistic as it only treats the set of counterterms associated with the intrinsic geometry of the AdS$_5$ submanifold.  However, since the background subtraction scheme leaves behind only divergences from the volume of the AdS$_5$ and we choose the boundary geometry to be $\mathds{S}^4\hookrightarrow\mathbb{R}^6$, only defect Weyl anomalies constructed purely from the intrinsic geometry should contribute, which will be accounted for in the scheme we have chosen. The caveat is that there may be structures for which we have not accounted in the full set of $11d$ counterterms, which is difficult to construct, whose pullback to the AdS$_5$ submanifold contains terms that contribute to the log divergence in a similar way. Absent a full holographic renormalization scheme for solutions to SUGRA dual to defects, which would replace background subtraction scheme as well, this scheme choice constructing counterterms only for the intrinsic geometry of the AdS$_5$ submanifold is the best tool available.

 Moving on, the volume of AdS$_5$ has well known divergences.  In order to systematically remove them and reveal any universal log-divergent terms, we consider AdS$_5$ in global coordinates with an $\mathds{S}^4$ boundary:
\begin{align}
    ds_{\rm AdS_5}^2 = dx^2 + \sinh^2(x)~d\Omega_4^2.
\end{align}
For simplicity, we consider the round metric on $\mathds{S}^4$.  Computing the AdS$_5$ volume requires regulating the large $x$ behavior, and so we introduce a radial cutoff $\Lambda_x\equiv-\log \frac{\eps_x}{2}$ for $\eps_x\ll 1$. Then, expanding in small $\eps_x$
\begin{align}
    \vol{\text{AdS}_5} = \frac{8\pi^2}{3}\int_0^{\Lambda_x}dx \sinh^4(x) = \frac{2\pi^2}{3\eps_x^4}-\frac{4\pi^2}{3\eps_x^2}-\pi^2\log\frac{\eps_x}{2}+\ldots~.
\end{align}
 We regulate the volume using covariant counterterms\footnote{To be complete, we should also fix finite counterterms to ensure that we are in a supersymmetry preserving scheme, but we will forego addressing this here as it is not germane to the problem at hand.} added on the radial cutoff slice that are standard in AdS$_5$ holographic renormalization  \cite{Henningson:1998gx,deHaro:2000vlm}
\begin{subequations}
    \begin{align}
    S_{\rm CT,1} &= -\frac{1}{4}\int d\Omega_4\sqrt{|g_{\eps_x}|} = -\frac{2\pi^2}{3\eps_x^4}+\frac{2\pi^2}{3\eps_x^2}-\frac{\pi^2}{4}+\ldots,\\
    S_{\rm CT,2} &= \frac{1}{48}\int d\Omega_4\sqrt{|g_{\eps_x}|}\CR_{\eps_x} = \frac{2\pi^2}{3\eps_x^2}-\frac{\pi^2}{3}+\ldots,
\end{align}
\end{subequations}
 where $\sqrt{|g_{\eps_x}|} =(1-\eps_x)^4\sqrt{|g_{\mathds{S}^4}|}/16\eps_x^2$ and $\CR_{\eps_x} =12\operatorname{csch}^2(\epsilon_x)$ are the volume form and the intrinsic Ricci scalar on the cutoff slice, respectively, built from the induced AdS$_5$ metric. Adding these counterterms to the bulk action, we see that the holographically renormalized volume of the unit AdS$_5$ takes the well-known form
\begin{align}\label{eq:Renormalized-AdS5-volume}
    \vol{\text{AdS}_5} =-\pi^2\log\frac{\eps_x}{2}+\ldots~.
\end{align} 
To complete the regularization of the on-shell actions for the vacuum AdS$_7\times\mathds{S}^4$ and two-charge solutions and extract the universal contributions to the defect free energy, we replace $\vol{\text{AdS}_5} =-\pi^2\log(\eps_x/2)$ wherever it appears.

\bibliographystyle{JHEP}
\bibliography{4d-Defect}
\end{document}